\documentclass[superscriptaddress,secnumarabic,nobibnotes,aps,prd,showkeys,showpacs,nofootinbib,onecolumn]{revtex4}
\usepackage{graphicx}
\usepackage{epsf}
\usepackage{bm}
\usepackage{amsmath}
\usepackage{amsfonts}
\usepackage{amssymb}
\usepackage{epstopdf}

\newcommand{\be}{\begin{equation}}
\newcommand{\ee}{\end{equation}}
\newcommand{\bear}{\begin{eqnarray}}
\newcommand{\eear}{\end{eqnarray}}

\newcommand{\mincir}{\raise
-3.truept\hbox{\rlap{\hbox{$\sim$}}\raise4.truept\hbox{$<$}\ }}
\newcommand{\magcir}{\raise
-3.truept\hbox{\rlap{\hbox{$\sim$}}\raise4.truept\hbox{$>$}\ }}

\input{tcilatex}
\begin{document}

\title{Dynamics in Varying vacuum Finsler-Randers Cosmology}
\author{G. Papagiannopoulos}
\email{yiannis.papayiannopoulos@gmail.com}
\affiliation{Faculty of Physics, Department of Astronomy-Astrophysics-Mechanics
University of Athens, Panepistemiopolis, Athens 157 83, Greece}
\author{S. Basilakos}
\email{svasil@academyofathens.gr}
\affiliation{Academy of Athens, Research Center for Astronomy and Applied Mathematics,
Soranou Efesiou 4, 11527, Athens, Greece}
\affiliation{National Observatory of Athens, Lofos Nymphon - Thissio, PO Box 20048 -
11810, Athens, Greece}
\author{A. Paliathanasis}
\email{anpaliat@phys.uoa.gr}
\affiliation{Instituto de Ciencias F\'{\i}sicas y Matem\'{a}ticas, Universidad Austral de
Chile, Valdivia, Chile}
\affiliation{Institute of Systems Science, Durban University of Technology, PO Box 1334,
Durban 4000, Republic of South Africa}
\author{Supriya Pan}
\email{supriya.maths@presiuniv.ac.in}
\affiliation{Department of Mathematics, Presidency University, 86/1 College Street,
Kolkata 700073, India}
\author{P. Stavrinos}
\email{pstavrin@math.uoa.gr}
\affiliation{Department of Mathematics, National and Kapodistrian University of Athens,
Panepistimiopolis 15784, Athens, Greece}

\begin{abstract}
In the context of Finsler-Randers theory we consider, for the first time,
the cosmological scenario of the varying vacuum. In particular, we assume
the existence of a cosmological fluid source described by an ideal fluid and
the varying vacuum terms. We determine the cosmological history of this
model by performing a detailed study on the dynamics of the field equations.
We determine the limit of General Relativity, while we find new eras in the
cosmological history provided by the geometrodynamical terms provided by the
Finsler-Randers theory.
\end{abstract}

\maketitle
\date{\today}

\section{Introduction}

Since the pioneering discovery of the accelerating expansion of our Universe 
\cite{Riess:1998cb,Perlmutter:1998np,Aghanim:2018eyx} cosmology is now in
the limelight of modern science. The physical mechanism able to explain this
accelerating universe is one of the greatest challenges of modern physics.
Within the realm of General Relativity (GR) this acceleration is easily
accommodated by introducing a dark energy sector(DE) \cite{Copeland:2006wr}
characterized by negative pressure. The simplest DE model arises with the
inclusion of a positive and time-independent cosmological constant, namely $%
\Lambda$, in the gravitational equations of GR \cite%
{Peebles:2002gy,Padmanabhan:2002ji}. The resulting cosmological scenario is
widely known as the $\Lambda$-cosmology and this cosmological model is in
agreement with a series of observational data, however it suffers from two
mayor problems, for details see \cite{Weinberg:1988cp}.

This naturally leads to think of several alternative $\Lambda$-cosmological
models \cite{Copeland:2006wr} to investigate the same issue. One of the
simplest and natural generalizations of the $\Lambda$-cosmology is to
introduce time dependence in the $\Lambda$ term, which leads to varying
vacuum cosmologies. On the other hand, apart from the concept of DE physics,
an alternative route to mimic this accelerating phase appears either due to
the direct modifications of GR leading directly to modified gravitational
theories \cite%
{Sotiriou:2008rp,DeFelice:2010aj,Clifton:2011jh,Capozziello:2011et,Nojiri:2017ncd}
or by introducing new gravitational theories completely different from GR,
such as the teleparallel equivalent of GR (TEGR) \cite{Cai:2015emx}.

The models arising from this latter approach are usually known as the
geometric dark energy (GDE) models. Although both DE and GDE models have
been widely studied and acknowledged in the literature, research over the
last several years has indicated that despite a large number of models, none
of them can be considered to be a completely healthy and viable model able
to portray the dynamical evolution of the universe. Most notably though, the
physical nature and evolution of both DE and GDE are still unknown even
after substantial cosmological research. Thus, the debates in search of a
perfect cosmological theory have been the central theme of modern cosmology
at present times. The studies so far clearly justify that there are
definitely no reasons to favor any particular cosmological theory or model,
at least in light of the recent cosmological observations.

An interesting gravitational theory in the context of the present
accelerating expansion is based on the introduction of Finsler geometry,
which gives rise to a wider geometrical picture of the universe extending
the traditional Riemannian geometry. In other words, one can recover the
Riemannian geometry as a special case of the Finslerian geometry. Thus
Finslerian geometry is expected to provide more insights on the dynamics and
evolution of the observed universe, and as a consequence, the cosmology in
Finslerian geometry gained significant attention in the scientific community
(eg \cite%
{Perelman,Minas,Ikeda,Kouretsis,Hohman,hohman2,Edwards,Vacaru,Caponio,Gibbons}%
). In particular, the Finsler-Randers (FR) metric \cite{Randers} and the
induced cosmological model \cite{Stav07,StavFluids} is of special interest
since the field equations include an extra geometrical term that acts as a
DE fluid. As we pointed out the (FR)\ cosmological model contains in each
point two metric structures, one Riemannian and one Finslerian so it can be
considered as a direction-dependent $(-y)$ motion of the Riemannian /FRW
model with osculating structure.

In the present article we consider a very general dynamical picture of the
universe in which a time-dependent cosmological term is present within the
context of Finslerian geometry. The presence of a time dependent
cosmological term, $\Lambda(t)$ actually inherits an interaction in the
cosmic sector. These kind of models are widely accepted in the literature
for their ability to describe various cosmological eras. The plan of the
paper is as follows.

In Section \ref{section2} we briefly discuss the FR cosmology.\ The varying
vacuum model is described in Section \ref{sec3} where we present the field
equations and the models of our analysis. Section \ref{sec4} includes the
main material of this work. In particular we present the dynamical analysis
and we determine the cosmological evolution for the models of our
consideration. Finally, in Section \ref{sec5} we discuss our results.

\section{Finsler-Randers theory: An overview}

\label{section2}

The origin of the FR model is based on the Finslerian geometry \cite{Stav07,
StavFluids} which is a natural generalization of the traditional Riemannian
geometry and it has gained considerable attention in the cosmological
community, see for instance \cite{15,Bekenstein,Mir,Bao,13,Gibbons}\emph{\ }%
for more details in this direction. In what follows we describe the basics
of the Finslerian geometry.

As shown by Asanov \cite{Asa41}, the general action for the osculating
Riemannian space-time of Einstein field equations is derived by a
variational principle of the integral action, $\displaystyle I_{_{G}}=\int L%
\big(x,y(x)\big)\sqrt{-g\big(x,y(x)\big)}dx^{4}$\ of an osculating
Riemannian procedure, where $L\big(x,y(x)\big)$\ is the osculating Ricci
scalar, in the context of Finsler Geometry. The derived field equations are
more general than the Riemannian ones \cite{Asa41}. These equations can also
be derived in the case of the Finsler-Randers model by making further
assumptions \cite{2}. Indicative works in the Finsler-Randers model are (%
\cite{4}-\cite{12},\cite{Randers}).

Given a differentiable manifold $M$, the Finsler space is generated from a
generating differentiable function $F(x,y)$ on the tangent bundle $TM$ with $%
F:\tilde{T}M\rightarrow R~,~\tilde{T}M=T(M)\backslash\{0\}$. The function $F$
is a one degree homogeneous function with respect to the variable $y$ which
is related to $x$, as $y=\frac{dx}{dt}$, here $t$ is the time variable. In
the FR space-time, we have 
\begin{equation*}
F(x,y)=\ \sigma(x,y)+v_{\mu}(x)y^{\mu},\;\;\;\sigma(x,y)=\sqrt{a_{\mu\nu
}y^{\mu}y^{\nu}},
\end{equation*}
where $a_{\mu\nu}$ is a Riemannian metric and $v_{\mu}=(v_{0},0,0,0)$ is a
weak primordial vector field with $\Vert v_{\mu}\Vert\ll1$. Let us note that
the vector field $v_{\mu}$ intrinsically contributes to the geometry of
Finslerian space-time and this vector field introduces a preferred direction
in the referred space time. The vector field $v_{\mu}$ additionally causes a
differentiation of geodesics from a Riemannian spacetime \cite{18}.
Although, there is a case where the geodesics of Riemannian and (FR) are
identical. This happens when the covector $v_{\mu}$ is a gradient vector.

In this formulation, in general one starts with the Lorentz symmetry
breaking, which is a common feature within quantum gravity phenomenology.
Such a departure from relativistic symmetries of space-time, leads to the
possibility for the underlying physical manifold to have a broader geometric
structure than the simple pseudo-Riemann geometry. 

\ One of the most characteristic features of Finsler geometry is the
dependence of the metric tensor to the position coordinates of the
base-manifold and to the tangent vector of a geodesic congruence, and this
velocity dependence reflects the Lorentz-violating character of the
kinematics.

The main object in Finsler geometry is the fundamental function $F(x,dx)$\
that generalizes the Riemannian notion of distance (\cite{Amendola:2003eq},%
\cite{Cai:2004dk},\cite{Pavon:2005yx}). In Riemann geometry the latter is a
quadratic function with respect to the infinitesimal increments $dx^{a}$\
between two neighboring points. Keeping all the postulates of Riemann
geometry but accepting a non-quadratic distance measure, a metric tensor can
be introduced as

\begin{equation}
f_{\mu \nu }=\ \frac{1}{2}\frac{\partial ^{2}F^{2}}{\partial y^{\mu
}\partial y^{\nu }}.  \label{fmn}
\end{equation}%
with tangent $y^{a}=\frac{dx^{a}}{d\tau }$. Note that when the generating
function $F(x,y)$\ is quadratic, the above definition is still valid and
leads to the metric tensor of Riemann geometry. The dependence of the metric
tensor to the position coordinates $x^{a}$\ and to the fiber coordinates $%
y^{a}\ $suggests that the geometry of Finsler spaces is a geometry on the
tangent bundle (TM). In other words, the Finsler manifold is a fiber space
where tensor fields depend on the position and on the infinitesimal
coordinate increments $y^{a}$. Therefore, the position dependence of Riemann
geometry can be replaced by the so called element of support, which is the
pair $(x^{a},y^{a})$. 

In relativistic applications of Finsler geometry the role of the supporting
direction $y^{a}$\ must be explicitly given. The locally anisotropic
character ($y$\ - dependent) of the gravitational field, can be appeared by
Lorentz violations, scalar / vector / spinor fields, or internal
perturbations in its structure. Energy momentum tensor of a cosmological
fluid in our consideration has the form

\begin{equation}
\displaystyle T_{_{\mu\nu}}\big(x,y(x)\big)=(\rho+P)y_{_{\mu}}(x)y_{_{%
\nu}}(x)-Pf_{_{\mu\nu}}\big(x,y(x)\big)
\end{equation}

It is a fundamental physical concept in the osculating Riemannian
(Finslerian) framework of Finsler gravity. By using an extending framework
of general relativity with a local anisotropic structure, the gravitational
field obtains more degrees of freedom. The geometrical concepts, as Ricci
tensors etc. in (\cite{Stav07}), are incorporated in the generalized
Friedmann equations, including the additional term $\dot{u}_{_{o}}$. This
represents the variation of small values of anisotropy. The consideration of
such a form of equations gives us the possibility of understanding of
possible small anisotropies of the evolution of the universe of the early
time up to the late time era. For cosmological applications of the Finsler
Randers models see \cite{4,5,6,7,12}.

From eq \textbf{(\ref{fmn})} one can now derive the Cartan tensor $C_{\mu\nu
k}=\frac{1}{2}\frac{\partial f_{\mu\nu}}{\partial y^{k}}$ using the
Finslerian metric tensor given above. We also note that the component $u_{0}$
can be given as $u_{0}=2C_{000}$ \cite{Stav07}. Let us consider the
gravitational equations in the FR\ cosmology in order to explore the
dynamics of the universe within this context. The field equations in this
context are 
\begin{equation}
L_{\mu\nu}=\ 8\pi G\left( T_{\mu\nu}-\frac{1}{2}Tf_{\mu\nu}\right) ,
\label{EE}
\end{equation}
where $L_{\mu\nu}$ denotes the Finslerian Ricci Tensor (for more details see 
\cite{Stav07});$~T_{\mu\nu}$ is the energy momentum tensor of the matter
sector and $T$ is the trace of $~T_{\mu\nu}$.

Now, consider the Finslerian perfect fluid with velocity 4-vector field $%
u_{\mu}$ for which the energy momentum tensor takes the form $T_{\mu\nu
}=diag\left( \rho,-Pf_{ij}\right) $, where $\{\mu,\nu\}\in\{0,1,2,3\}$ and $%
\{i,j\}\in\{1,2,3\}$; $\rho$ and $P$ respectively denote the total energy
density and pressure of the underlying cosmic fluid (\cite{Asa41}).

For the above expression of the energy-momentum tensor, in a spatially flat
Friedmann-Lema\^{\i}tre-Robertson-Walker (FLRW) metric\footnote{%
Let us note that the nonzero components of the Ricci tensor in the context
are: $L_{00}=3(\frac{\ddot{a}}{a}+3\frac{\dot{a}}{4a}\dot{u}_{0})$ and $%
L_{ii}=-(a\ddot{a}+2\dot{a}^{2}+\frac{11}{4}a\dot{a}\dot{u}_{0})/\Delta _{ii}
$ where $(\Delta _{11},\Delta _{22},\Delta _{33})=(1,r^{2},r^{2}\mathrm{sin}%
^{2}\theta )$.}, 
\begin{equation*}
ds^{2}=-dt^{2}+a^{2}\left( t\right) \left( dx^{2}+dy^{2}+dz^{2}\right) ,
\end{equation*}%
the gravitational field equations can be explicitly written as \cite{Stav07} 
\begin{equation}
\dot{H}+H^{2}+\frac{3}{4}HZ_{t}=-\frac{4\pi G}{3}(\rho +3p),  \label{e1}
\end{equation}%
\begin{equation}
\dot{H}+3H^{2}+\frac{11}{4}HZ_{t}=4\pi G(\rho -p),  \label{e2}
\end{equation}%
where and the overdot represents the derivative with respect to the cosmic
time and $H\equiv \dot{a}/a$, is the Hubble rate and $Z_{t}=\dot{u}_{0}(t)$.
Now, combining Eqs. (\ref{e1}) and (\ref{e2}) one arrives at 
\begin{equation}
H^{2}+HZ_{t}=\frac{8\pi G}{3}\rho .  \label{6}
\end{equation}%
Obviously, the Friedmann equations are modified by the extra term $HZ_{t}$.
As expected for $Z_{t}=0$, hence $u_{0}\equiv 0$ we recover the usual
Friedmann equations.


Additionally, using the Bianchi identities one can have the conservation
equation for the total fluid which goes as 
\begin{equation}
\dot{\rho}+3H\left( \rho +p\right) -Z_{t}\left( \rho +\frac{3}{2}p\right) =0.
\label{e3}
\end{equation}%
This clearly shows that the usual conservation equation of the
energy-momentum tensor does not hold in the FR geometry. This consequently
means that the FR geometry is naturally endowed with the effective matter
creation process which is quantified through the extra geometrodynamical
term appearing in eqn. (\ref{e3}), namely, $Z_{t}(\rho +\frac{3}{2}p)$. 

\bigskip Observing the form of the above conservation equation for the CDM
case $(p=0)$, and comparing to the creation of cold dark matter model \cite%
{mat1},\cite{mat2},\cite{mat3}, we deduce that in the scenario at hand we
obtain an effective matter creation model of (modified) gravitational
origin. In particular, based on the aforementioned articles one can define
the dark matter density by the following equation $\dot{\rho}+3H\rho =\Gamma
\rho $, while the the creation pressure of CDM component is given by $p_{c}=-%
\frac{\Gamma \rho }{3H}$. Notice that $\Gamma $\ is the creation rate of CDM
particles (see \cite{prigogine},\cite{calvao},\cite{matrest}). Therefore
combining the above equation with (\ref{e3}) the effective matter creation
rate is written in terms of $Z_{t}$\ which is a geometrical quantity, namely 
$\Gamma =-Z_{t}$and thus the effective creation pressure reads $p_{c}=\frac{%
\rho Z_{t}}{3H}$. The particle production is an irreversible process, and,
as such, it should be constrained by the second law of thermodynamics. A
possible macroscopic solution for this problem was discussed by Prigogine and%
\textbf{\ }collaborators \cite{prigogine} utilizing nonequilibrium
thermodynamics for open systems, and by Calvao, Lima \& Waga \cite{calvao}
through a covariant relativistic treatment for imperfect fluids (see also 
\cite{limaGermano}). In this framework particle production, at the expense
of the gravitational field, is an irreversible process constrained by the
usual requirements of nonequilibrium thermodynamics. This irreversible
process is described by a negative pressure term in the stress tensor whose
form is constrained by the second law of thermodynamics. It is interesting
to mention that the proposed macroscopic approach has also microscopically
been justified by Zimdahl and collaborators via a relativistic kinetic
theoretical formulation (see \cite{zim1},\cite{zim2}). In comparison to the
standard equilibrium equations, the irreversible creation process is
described by two new ingredients: a balance equation for the particle number
density and a negative pressure term in the stress tensor. These quantities
are connected to each other in a very definite way by the second law of
thermodynamics. 

In general the idea of cosmological particle production or matter creation
was discussed extensively indepedently by several authors \cite{Parker1968,
Parker1969, Parker1970, Parker1977},\cite{ZS1972, ZS1977},\cite%
{LimaBasilakos},\cite{Grib1974, Grib1976, Grib1994} where they proposed that
the gravitational field of the expanding universe is constantly acting on
the quantum vacuum, and due to this, particles are created. This creation
process is a continuous phenomenon and the created acquire their mass,
momentum and energy. Over the last decade, cosmological theories with matter
creation, have been extensively studied and different models have been
constrained in presence of the observational datasets (see \cite%
{Steigman:2008bc,Lima:2009ic,Nunes:2015rea,Pigozzo:2015swa,Nunes:2016aup}
and the references therein). In particular, a recent analysis \cite%
{Pigozzo:2015swa} has argued that the creation of dark matter particles is
favored (within 95\% CL) according to the available observational sources,
however, it is important to mention that rate of matter creation must be
constrained in such a way so that the matter sector does not deviate much
from its standard evolution ($\propto a(t)^{-3}$), see \cite{Nunes:2015rea}
for details. 

\section{Varying Vacuum in a Finsler Randers Model}

\label{sec3} In the framework of General Relativity the Running Vacuum Model
(RVM) has been thoroughly studied at the background and perturbation levels
respectively (see \cite{VV} and references therein). Here we want to extend
the situation by including in the Finsler Randers geometry the concept of
RVM. Notice that the time dependence of the vacuum energy density in the RVM
is only through the Hubble rate, hence ${\dot{\rho}}_{\Lambda}\neq0$.

Let us assume that we have a mixture of two fluids, namely, matter (labeled
with the symbol $m$) and the varying vacuum (labeled with $\Lambda$), hence
the total energy density and pressure of the total fluid are given by 
\begin{equation}
\rho=\rho_{m}+\rho_{\Lambda},\;\;\;\;\;p=p_{m}+p_{\Lambda}.  \label{Eq1}
\end{equation}

The complete set of field equations reads

\begin{equation}
1+\frac{Z_{t}}{H}=\frac{8\pi G}{3H^{2}}\rho_{m}+\frac{8\pi G}{3H^{2}}%
\rho_{\Lambda}  \label{fr11}
\end{equation}%
\begin{equation}
\frac{\dot{H}}{H^{2}}+1+\frac{3Z_{t}}{4H}=-\frac{4\pi G}{3H^{2}}(\rho
_{m}+3w_{m}\rho_{m}-2\rho_{\Lambda}),  \label{fr2}
\end{equation}
where through the Bianchi equations the continuity equations become,

\begin{align}
\dot{\rho}_{m}+3H(1+w_{m})\rho_{m}-Z_{t}\left( \rho+\frac{3}{2}p\right) & =%
\hat{Q},  \label{bianchi1} \\
\dot{\rho}_{\Lambda} & =-\hat{Q},  \label{Eq31}
\end{align}
where $w_{m}=p_{m}/\rho_{m}$, is the equation-of-state parameter of the
matter fluid and the term $\hat{Q}$ appearing in (\ref{bianchi1}) and (\ref%
{Eq3}) refers to the interaction rate between the matter and vacuum sectors.
As one can quickly note that $\hat{Q}=0$ actually recovers the usual
non-interacting dynamics. It is easy to realize that the presence of
interaction between these sectors certainly generalizes the cosmic dynamics
and it is of utmost importance to address many cosmological puzzles. Due to
the diverse characteristics, interacting models have gained a massive
attention to the cosmological community because. The mechanism of an
interaction in the dark sector of the universe is a potential route that may
explain the cosmic coincidence problem \cite%
{Amendola:1999er,Amendola:2003eq,Cai:2004dk,Pavon:2005yx,delCampo:2008sr,delCampo:2008jx}
and provide a varying cosmological constant that could explain the tiny
value of the cosmological constant leading to a possible solution to the
cosmological constant problem \cite{Wetterich:1994bg}. In the past years, a
cluster of interaction models have been studied by many researchers. Some of
the interaction models existing in the literature are \cite%
{Barrow:2006hia,Amendola:2006dg,Pavon:2007gt,Chimento:2009hj,Arevalo:2011hh,Yang:2018xlt,an001,Tsiapi:2018she,Yang:2018qec,Pan:2020mst}
while some cosmological constraints on interacting models can be found in 
\cite%
{Salvateli,QproptorhoL,Nunes:2016dlj,Pan:2016ngu,Sharov:2017iue,Yang:2017yme,Yang:2017ccc,Yang:2017zjs,Pan:2017ent,Yang:2018euj,in1,in2,in3,Pan:2019jqh,Pan:2019gop,in4,in5,DiValentino:2019ffd,DiValentino:2019jae,Pan:2020zza}%
. On the other hand, this model can also be seen as the particle creation
model which has gained massive attention in the scientific society \cite%
{Prigogine-inf,Abramo:1996ip,Gunzig:1997tk,Zimdahl:1999tn,RCN16,deHaro:2015hdp, sp1, Nunes:2016aup,sp2}%
. In this work we aim to study the generic evolution of the solution of the
field equations for specific functional forms of the interaction term $\hat{Q%
}$. In the following we replace the interaction term $\hat{Q}$ with $Q=\hat{Q%
}+Z_{t}\left( \rho+\frac{3}{2}p\right) $ such that to remove the nonlinear
term and rewrite the continuous equation in the friendly form%
\begin{align}
\dot{\rho}_{m}+3H(1+w_{m})\rho_{m} & =Q,  \label{bianchi} \\
\dot{\rho}_{\Lambda} & =-Q,  \label{Eq3}
\end{align}

Following our previous works \cite{VarVacGP,finslerGP} we study how the
implementation of the Finsler geometry affects the varying vacuum scenarios
studied in GR as well as how the implementation of the varying vacuum
responds in a Finsler Randers scenario.

\section{Dynamical evolution}

\label{sec4}

In this Section, we study the cosmological evolution of the different
cosmological scenarios of varying vacuum in a FR geometrical background by
using methods\ of the dynamical analysis \cite{con1,con2}. Specifically, we
study the critical points of the field equations in order to identify the
different cosmological eras that are accommodated by each scenario. The
respective stability of these cosmological eras is determined by calculating
the eigenvalues of the linearized system at the specific critical points.

In order to perform such an analysis we define proper dimensionless
variables such that to rewrite the field equations as a set of
algebraic-differential equations. The critical points of the system are
considered to be the sets of variables for which all the ODE of the system
are zero. These sets of variables correspond to a specific solution of the
system and each to a different era of the cosmos that may be able to
describe the observed universe. The eigenvalues of the above points are
defining the stability of the critical points. Namely a critical point is
stable/attractor when the corresponding eigenvalues have negative real
parts. Thus, the eigenvalues are valuable tools that characterize the
behavior of the dynamical system around the critical point \cite{wiggins}.

Our approach is as follows. We consider a dynamical system of any dimension 
\begin{equation*}
\dot{x}^{A}=f^{A}\left( x^{B}\right) ,
\end{equation*}
and then a critical point of the system $P=P\left( x^{B}\right) $ which has
to satisfy \ $f^{A}\left( P\right) =0$. The linearized system around $P$ is
written as%
\begin{equation*}
\delta\dot{x}^{A}=J_{B}^{A}\delta x^{B},~J_{B}^{A}=\frac{\partial
f^{A}\left( P\right) }{\partial x^{B}}.
\end{equation*}
where $J_{B}^{A}$ is the respective Jacobian matrix. We calculate the
eigenvalues and eigenvectors and then express the general solution at the
respective points as their expression. Since the linearized solutions are
expressed in terms of the eigenvalues $\lambda_{i}$ and thus as functions of 
$e^{\lambda_{i}t}$, apparently when all these terms have their real part
negative the respective solution of the critical point is stable and the
point is an attractor, otherwise the point is a source.

Such an analysis is very useful in terms of defining viable theories that
can describe the observable universe. Thus for a healthy theory to be
viable, the critical point analysis should provide points where the universe
will be accelerating and also these points to be stable. This analysis has
been applied in various cosmological models, for instance see and references
therein \cite%
{VarVacGP,finslerGP,Quintessence,aetherGP,BransDickeGP,dyn2,dyn3,dyn4,dyn5,dyn6,dyn7,dyn8,dyn9,pano1,Kerachian}%
.

\subsection{Dimensionless variables}

We select to work in the $H-$normalization. Therefore, we define the
dimensionless variables \cite{con1,con2}

\begin{equation}
\Omega
_{m}=\frac{\rho_{_{m}}}{3H^{2}},~%
\Omega
_{z}=\frac{Z_{t}}{H},~%
\Omega
_{\Lambda}=\frac{\rho_{\Lambda}}{3H^{2}}.  \label{dimensionless}
\end{equation}

Thus, the first Friedmann equation gives the constraint equation%
\begin{equation}
1+\Omega_{z}=\Omega_{m}+\Omega_{\Lambda}\   \label{con}
\end{equation}
while the rest of the field equations are written as follows%
\begin{equation}
\frac{d%
\Omega
_{\Lambda}}{d\ln a}=2\Omega_{\Lambda}\left[ 1+\frac{3}{4}(\Omega_{m}+%
\Omega_{\Lambda}\ -1)+\frac{1}{2}\Omega_{m}(1+3w_{m})-\Omega_{\Lambda }%
\right] -\frac{Q}{3H^{3}},  \label{ds1}
\end{equation}

\begin{equation*}
\frac{d%
\Omega
_{m}}{d\ln a}=2\Omega_{m}\Bigg[1+\frac{3}{4}(\Omega_{m}+\Omega_{\Lambda }\
-1)+\frac{1}{2}\Omega_{m}(1+3w_{m})-\Omega_{\Lambda}\Bigg]+\frac{Q}{3H^{3}}%
-3\Omega_{m}(1+w_{m}),
\end{equation*}

where $p_{m}=w_{m}\rho_{m}$. In the following we assume that $w_{m}\in\left(
-1,1\right) $.

We proceed by determining the critical points of the dynamical system. Every
point $P$ has coordinates $P=\mathbf{\{}%
\Omega
_{m},%
\Omega
_{\Lambda},%
\Omega
_{z}\mathbf{\},}$ and describes a specific cosmological solution. For every
point we determine the physical cosmological variables as well as the
equation of the state parameter $w_{tot}\left( P\right) $. In order to
determine the stability of each critical point the eigenvalues of the
linearized system around the critical point $P$ are derived.

We remark that the second Friedmann equation with the use of the
dimensionless variables reads

\begin{equation}
\frac{\dot{H}}{H^{2}}=-1-\frac{3}{4}\Omega_{z}-\frac{1}{2}%
\Omega_{m}(1+3w_{m})\ +\Omega_{\Lambda}\ ,  \label{accel}
\end{equation}
from where we find that at a stationary point $P$, the equation of state
parameter for the effective fluid is 
\begin{equation}
w_{tot}\left( P\right) =-\frac{1}{3}+\frac{2}{3}\left( \frac{3}{4}\Omega_{z}+%
\frac{1}{2}\Omega_{m}(1+3w_{m})-\Omega_{\Lambda}\right) .
\end{equation}

In this work we study various functional forms for the interaction term $Q$.
In order to extend the results of \cite{VarVacGP}, we assume that (A) the
interaction term $Q$ is proportional to the density of dark matter \cite%
{Salvateli}, that is, $Q_{A}=9nH\rho_{m}$ or equivalent $Q_{A}\simeq9nH^{3}%
\Omega_{m},$ where the dimensionless parameter $n$ is an indicator of the
interaction strength; (B) $Q$ is proportional to the density of the dark
energy term, i.e. $Q_{B}=9nH^{3}\rho_{\Lambda}~$\cite%
{Tsiapi:2018she,QproptorhoL}; (C) $\ Q_{C}$ is proportional to the sum of
the energy density of the dark sector of the universe, that gives $%
Q_{C}=9nH(\rho_{\Lambda}+\rho_{m})$.

Motivated by the above functional forms $Q$, which have given interesting
results in General Relativity, \cite{VarVacGP}, we propose some new
interaction terms which are proportional to the energy density $\Omega_{z}$.
In particular we select the models\ (D) $Q_{D}=9nH^{3}\Omega_{z};~$(E) $%
Q_{E}=9nH^{3}\Omega_{z}+9mH\rho_{m}$ $\ $and (D)~$Q_{F}=9nH^{3}\Omega
_{z}+9mH\rho_{m}$. In these models $m$ is a dimensionless parameter, an
indicator of the interaction strength. Finally, in order to compare our
results with the non-varying vacuum model we investigate the case where $%
Q_{G}=-3\Omega_{z}\Omega_{m}H^{3}\left( 1+\frac{3}{2}w_{m}\right) $.

\subsection{Model A: $Q_{A}=9nH\protect\rho_{m}$}

For the first model that we consider $Q_{A}=9nH\rho_{m}$, the field
equations are expressed as follows.

\begin{equation}
\frac{d%
\Omega
_{\Lambda}}{d\ln a}=2\Omega_{\Lambda}[1+\frac{3}{4}(\Omega_{m}+\Omega
_{\Lambda}\ -1)+\frac{1}{2}\Omega_{m}(1+3w_{m})-\Omega_{\Lambda}]-3n%
\Omega_{m}  \label{a1}
\end{equation}

\begin{equation}
\frac{d%
\Omega
_{m}}{d\ln a}=2\Omega_{m}(1+\frac{3}{4}(\Omega_{m}+\Omega_{\Lambda}\ -1)+%
\frac{1}{2}\Omega_{m}(1+3w_{m})-\Omega_{\Lambda})+3n\Omega_{m}-3%
\Omega_{m}(1+w_{m})  \label{a2}
\end{equation}

The dynamical system (\ref{a1}), (\ref{a2}) admits three critical points
with coordinates $\{%
\Omega
_{m},%
\Omega
_{\Lambda},%
\Omega
_{z}\}$%
\begin{equation*}
A_{1}=\{0,0,-1\},~A_{2}=\{0,1,0\},~A_{3}=\{1-\frac{n}{1+w_{m}},\frac {n}{%
1+w_{m}},0\},~
\end{equation*}

Point $A_{1}$ always exists and describes an empty universe with equation of
state parameter~$w_{tot}\left( A_{1}\right) =-\frac{5}{6}.$ The universe
accelerates with the contribution of the extra term introduced due to the
Finsler-Randers Geometry. The eigenvalues of the linearized system near to
point $A_{1}$ are $\{\frac{1}{2},-\ \frac{5}{2}-3(w_{m}-n))\}$, from where
we can infer that the point is always a source, since one of the eigenvalues
is always positive.

Point $A_{2}~$describes a de Sitter universe with equation of state
parameter~$w_{tot}\left( A_{2}\right) =-1$, where only the $\Lambda$\ term
contributes in the evolution of the universe. The eigenvalues are derived to
be $\left\{ -\frac{1}{2},-3(w_{m}-n+1)\right\} ,$ from where we can infer
that the point is an attractor when $w_{m}\geq n-1$ or equivalently $%
n\leq1+w_{m}$. Because $n$ is the strength of the interaction of the varying
vacuum and matter we assume this term to be close to zero (either positive
or negative) and thus understand that the aforementioned condition is
satisfied (we generally have that $w_{m}>-1$). Thus this point is of great
physical interest.

Point $A_{3}$ exists for $n\geq0~$(for $n<0$ then $w_{m}<-1$ and it exists
in the phantom region) and describes a universe dominated by the varying
vacuum and the matter fluid; in the case where $w_{m}=0$, point $A_{3}$
describes the $\Lambda$-CDM universe in the FR theory. The equation of state
parameter is derived $w_{tot}\left( A_{3}\right) =w_{m}-n$, from where we
conclude that the exact solution at the point describes an accelerated
universe when $w_{m}\leq n+\frac{1}{3}$. The eigenvalues of the linearized
system are\ $\left\{ 3(w_{m}-n+1),3(w_{m}-n+\frac{5}{6})\right\} $ and thus
can be stable for $w_{m}<n-1.$

The above results are summarized in Tables \ref{taba01} and \ref{taba02}. In
addition in the Figs. \ref{fig:fig1},\ref{fig:fig3} we present the evolution
of the trajectories for the dynamical system of our study.

\begin{figure}[ptb]
\centering
\includegraphics[width=0.40\textwidth]{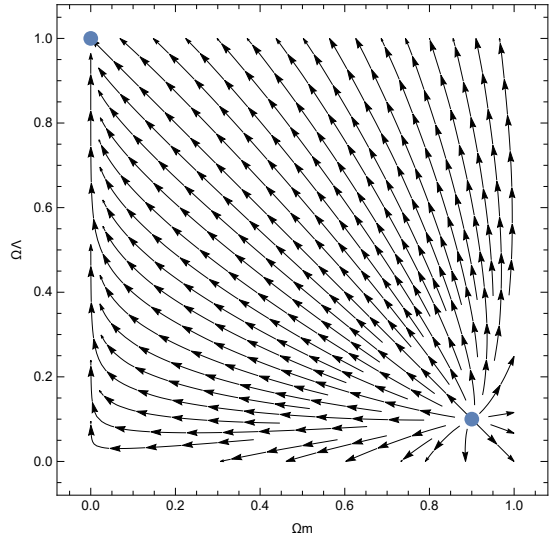}
\caption{Phase space diagram for the dynamical system (\protect\ref{a1}), (%
\protect\ref{a2}). We consider $w_{m}=0$, for $n<1$. The unique attractor is
the de Sitter point $A_{2}$. }
\label{fig:fig1}
\end{figure}
\begin{figure}[ptb]
\centering
\includegraphics[width=0.7\textwidth]{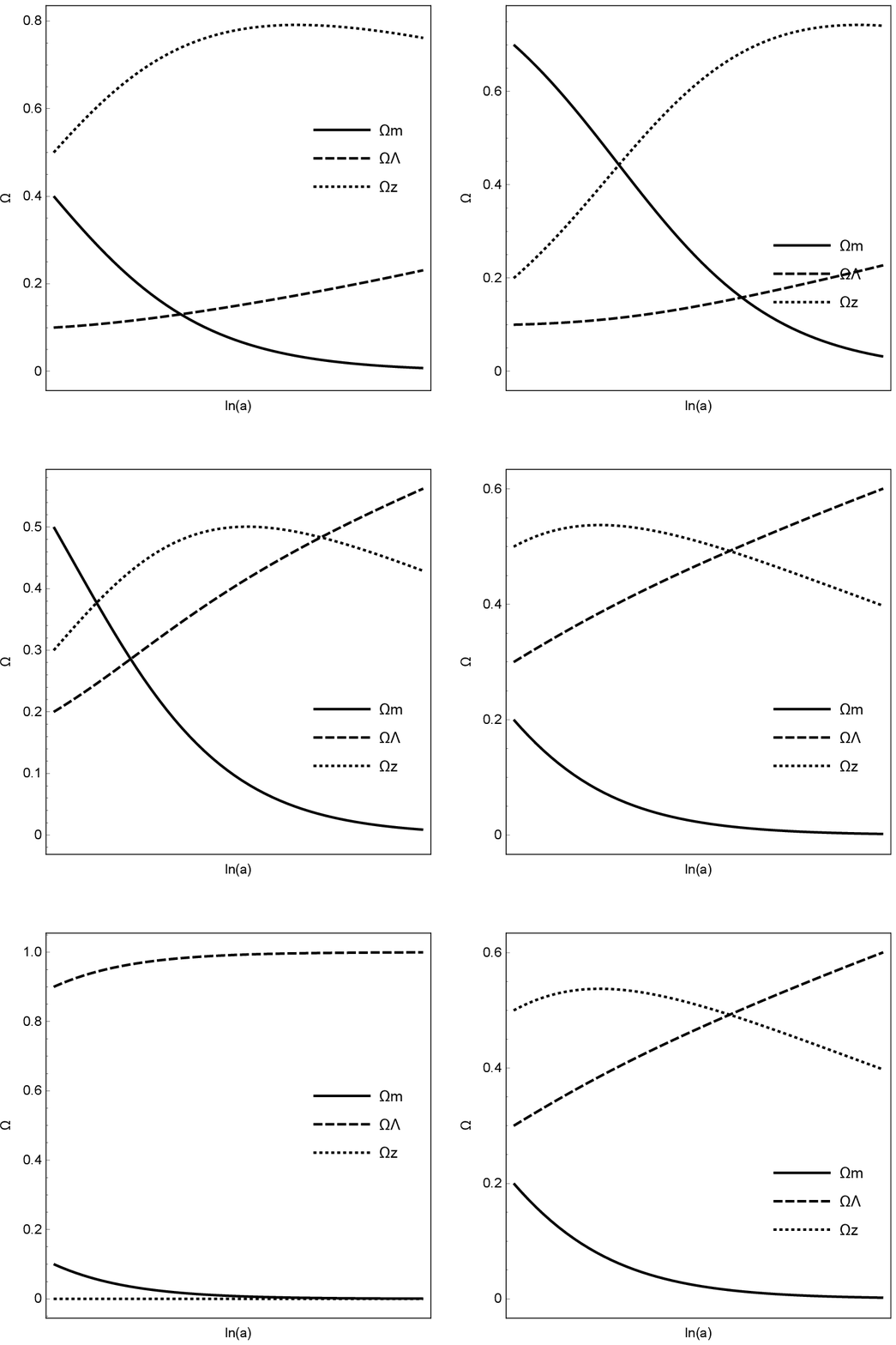}
\caption{Evolution diagrams with time, for various energy densities of the
dynamical system (\protect\ref{a1}), (\protect\ref{a2}). We consider the
initial conditions (a) $\Omega_{m}=0.4,$ $\Omega_{\Lambda}=0.1$ (b) $%
\Omega_{m}=0.7$, $\Omega_{\Lambda}=0.1$ (c) $\Omega_{m}=0.5$, $%
\Omega_{\Lambda}=0.2$ (d) $\Omega_{m}=0.2$, $\Omega _{\Lambda}=0.3$ (e) $%
\Omega_{m}=0.1$, $\Omega_{\Lambda}=0.9$ (f) $\Omega _{m}=0.2$, $%
\Omega_{\Lambda}=0.3$, for $n<1$ and $w_{m}=0$.}
\label{fig:fig3}
\end{figure}

\begin{table}[tbp] \centering%
\caption{Stationary points and physical parameters for the interaction model
A. }%
\begin{tabular}{ccccc}
\hline\hline
\textbf{Point} & $\left( \mathbf{%
\Omega
}_{m}\mathbf{,%
\Omega
}_{\Lambda}\mathbf{,%
\Omega
}_{z}\right) $ & \textbf{Existence} & $\mathbf{w}_{tot}$ & \textbf{%
Acceleration} \\ \hline
$A_{1}$ & $\left( 0,0,-1\right) $ & Always & $-\frac{5}{6}$ & Yes \\ 
$A_{2}$ & $\left( 0,1,0\right) $ & Always & $-1$ & Yes \\ 
$A_{3}$ & $\left( 1-\frac{n}{1+w_{m}},\frac{n}{1+w_{m}},0\right) $ & $%
w_{m}\neq-1,(n=0,w_{m}>-1)\ $or$~(n>0,w_{m}>-1+n)$ & $w_{m}-n$ & $w_{m}\leq
n-\frac{1}{3}$ \\ \hline\hline
\end{tabular}
\label{taba01}%
\end{table}%

\begin{table}[tbp] \centering%
\caption{Stationary points and stability conditions for the interaction
model A. }%
\begin{tabular}{ccc}
\hline\hline
\textbf{Point} & \textbf{Ei}$\text{\textbf{genvalues}}$ & \textbf{Stability}
\\ \hline
$A_{1}$ & $\{\frac{1}{2},-\ \frac{5}{2}-3(w_{m}-n))\}$ & Source \\ 
$A_{2}$ & $\left\{ -\frac{1}{2},-3(w_{m}-n+1)\right\} $ & $w_{m}\geq n-1$ \\ 
$A_{3}$ & $\left\{ 3(w_{m}-n+1),3(w_{m}-n+\frac{5}{6})\right\} $ & $%
w_{m}<n-1 $ \\ \hline\hline
\end{tabular}
\label{taba02}%
\end{table}%

\subsection{Model B: $Q_{B}=9nH\protect\rho_{\Lambda}$}

For the second model of our analysis, where $Q_{B}=9nH\rho_{\Lambda}$, the
field equations become

\begin{equation}
\frac{d%
\Omega
_{\Lambda}}{d\ln a}=2\Omega_{\Lambda}\left[ 1+\frac{3}{4}(\Omega_{m}+%
\Omega_{\Lambda}\ -1)+\frac{1}{2}\Omega_{m}(1+3w_{m})-\Omega_{\Lambda }%
\right] -3n\Omega_{\Lambda},  \label{b1}
\end{equation}
\begin{equation}
\frac{d%
\Omega
_{m}}{d\ln a}=2\Omega_{m}\left( 1+\frac{3}{4}(\Omega_{m}+\Omega_{\Lambda }\
-1)+\frac{1}{2}\Omega_{m}(1+3w_{m})-\Omega_{\Lambda}\right) +3n\Omega
_{\Lambda}-3\Omega_{m}(1+w_{m}),  \label{b2}
\end{equation}

The dynamical system (\ref{b1}), (\ref{b2}),admits three critical points
with coordinates 
\begin{equation*}
B_{1}=\{0,0,-1\},~B_{2}=\{1,0,0\},B_{3}=\left\{ \frac{n}{1+w_{m}},1-\frac {n%
}{1+w_{m}},0\right\} .
\end{equation*}

Point $B_{1}$ exists always and it corresponds to an empty universe with
equation of state parameter~$w_{tot}\left( B_{1}\right) =-\frac{5}{6}$, that
is accelerating due to the contribution of the extra term introduced by the
Finsler-Randers geometrical background. The eigenvalues of the linearized
system are $\{\frac{(1-6n)}{2},-\ \frac{(5+6w_{m})}{2}\};$ hence the exact
solution at the stationary point $B_{1}$ it is stable when $n>\frac{1}{6}~$%
and$~w_{m}>-\frac{5}{6}$. Thus this point is of great physical interest
since it can describe a past or future acceleration phase.

Point $B_{2}~$describes a universe dominated by matter, $w_{tot}\left(
B_{2}\right) =w_{m}$, and the exact solution at the point corresponds to an
accelerated universe when $w_{m}<-\frac{1}{3}$. The eigenvalues of the
linearized system are derived to be $\left\{ 3(w_{m}-n+1),\frac{(5+6w_{m})}{2%
}\right\} ,$ from where we observe that $B_{3}$ is an attractor when $n\leq%
\frac{1}{6}$\& $w_{m}<n-1\ $or$\ n>\frac{1}{6}~\&~w_{m}<-\frac{5}{6}$.

Point $B_{3}$ exists when $~w_{m}\neq-1,(1+w_{m})\geq n\geq0\ $and it has
the same physical properties with point $A_{3}$. The eigenvalues of the
linearized system near the stationary point are derived to be $\left\{ \frac{%
(6n-1)}{2},-3(w_{m}-n+1)\right\} $, from where we infer that the exact
solution at $B_{3}$ is stable for $n<\frac{1}{6}~\&~w_{m}>-1+n$.

\begin{figure}[ptb]
\centering\includegraphics[width=0.40\textwidth]{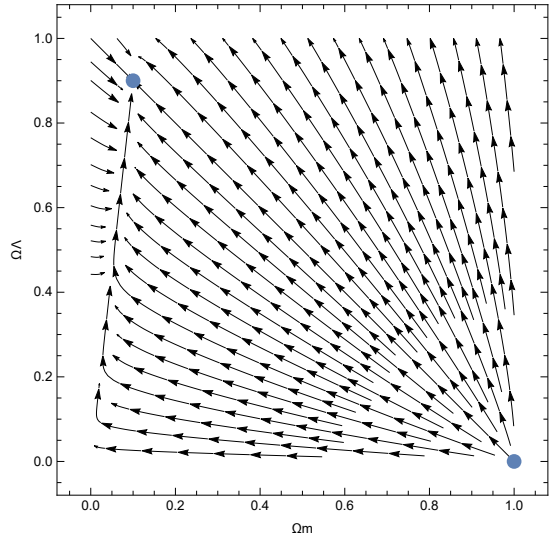}
\caption{Phase space diagram for the dynamical system (\protect\ref{b1}), (%
\protect\ref{b2}). We consider $w_{m}=0$, for $n<1$. The unique attractor is
the de Sitter point $B_{3}$. }
\label{fig:fig5}
\end{figure}

\begin{figure}[ptb]
\centering\includegraphics[width=0.7\textwidth]{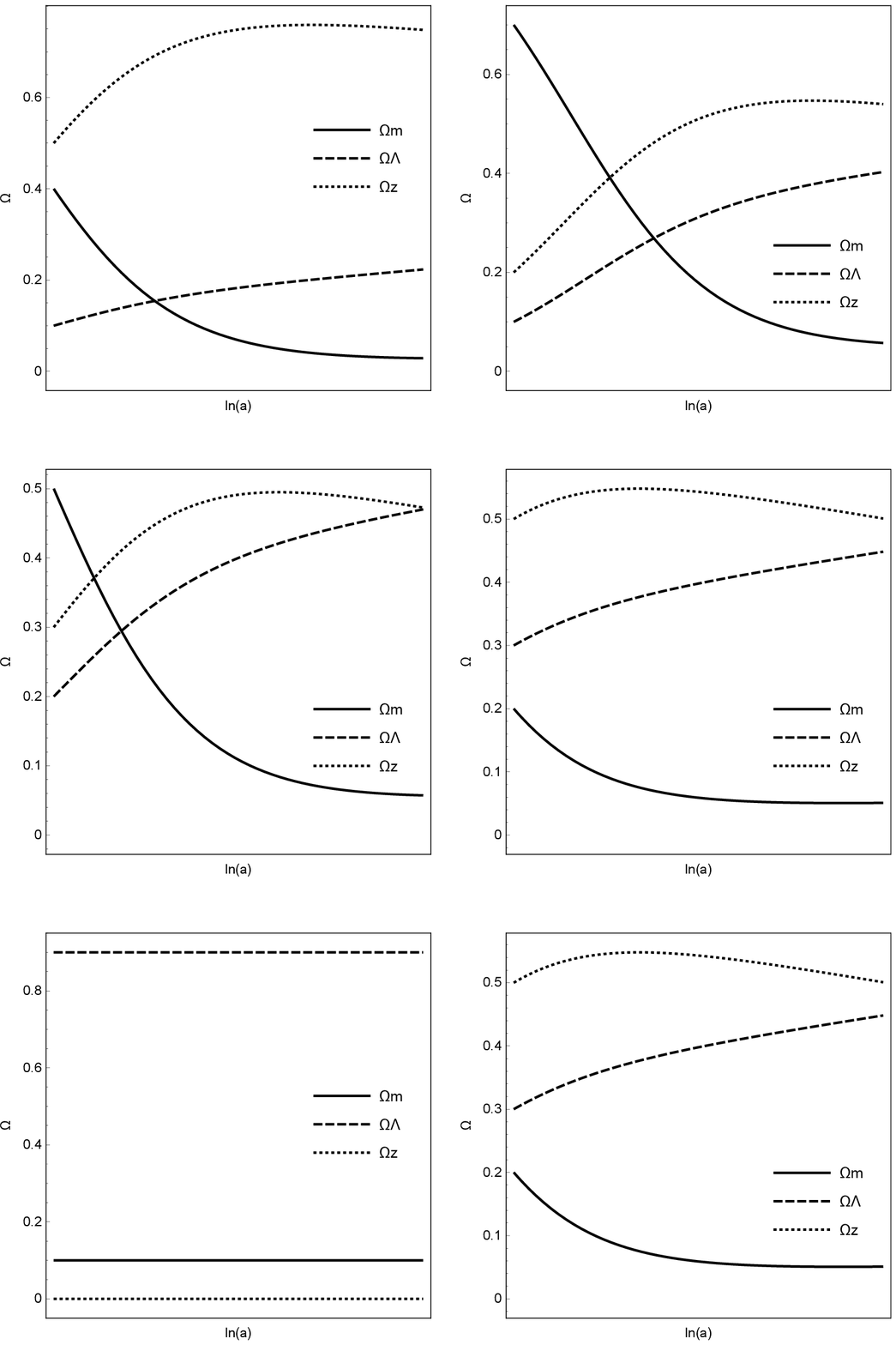}
\caption{Evolution diagrams with time, for various energy densities of the
dynamical system (\protect\ref{b1}), (\protect\ref{b2}). We consider the
initial conditions (a) $\Omega _{m}=0.4,$ $\Omega_{\Lambda}=0.1$ (b) $%
\Omega_{m}=0.7$, $\Omega_{\Lambda}=0.1$ (c) $\Omega_{m}=0.5$, $%
\Omega_{\Lambda}=0.2$ (d) $\Omega_{m}=0.2$, $\Omega_{\Lambda}=0.3$ (e) $%
\Omega_{m}=0.1$, $\Omega_{\Lambda}=0.9$ (f) $\Omega_{m}=0.2$, $%
\Omega_{\Lambda}=0.3$, for $n<1$ and $w_{m}=0$.}
\label{fig:fig7}
\end{figure}

The above results are summarized in Tables \ref{tabb01} and \ref{tabb02}. In
Figs. \ref{fig:fig5},~\ref{fig:fig7} the evolution of trajectories for the
dynamical system our study in phase space are presented.

\begin{table}[tbp] \centering%
\caption{Stationary points and physical parameters for the interaction model
B. }%
\begin{tabular}{ccccc}
\hline\hline
\textbf{Point} & $\left( \mathbf{%
\Omega
}_{m}\mathbf{,%
\Omega
}_{\Lambda}\mathbf{,%
\Omega
}_{z}\right) $ & \textbf{Existence} & $\mathbf{w}_{tot}$ & \textbf{%
Acceleration} \\ \hline
$B_{1}$ & $\left( 0,0,-1\right) $ & Always & $-\frac{5}{6}$ & Yes \\ 
$B_{2}$ & $\left( 1,0,0\right) $ & Always & $w_{m}$ & $w_{m}\leq-\frac{1}{3}$
\\ 
$B_{3}$ & $\left( \frac{n}{1+w_{m}},1-\frac{n}{1+w_{m}},0\right) $ & $%
w_{m}\neq-1,(1+w_{m})\geq n\geq0$ & $-1+n$ & $n\leq\frac{2}{3}$ \\ 
\hline\hline
\end{tabular}
\label{tabb01}%
\end{table}%

\begin{table}[tbp] \centering%
\caption{Stationary points and stability conditions for the interaction
model B. }%
\begin{tabular}{ccc}
\hline\hline
\textbf{Point} & \textbf{Ei}$\text{\textbf{genvalues}}$ & \textbf{Stability}
\\ \hline\hline
$B_{1}$ & $\{\frac{(1-6n)}{2},-\ \frac{(5+6w_{m})}{2}\}$ & $n>\frac{1}{6}%
~\&~w_{m}>-\frac{5}{6}$ \\ 
$B_{2}$ & $\left\{ 3(w_{m}-n+1),\frac{(5+6w_{m})}{2}\right\} $ & $n\leq 
\frac{1}{6}$\& $w_{m}<n-1\ $or$\ n>\frac{1}{6}~\&~w_{m}<-\frac{5}{6}$ \\ 
$B_{3}$ & $\{\frac{(6n-1)}{2},-3(w_{m}-n+1)\}$ & $n<\frac{1}{6}%
~\&~w_{m}>-1+n $ \\ \hline\hline
\end{tabular}
\label{tabb02}%
\end{table}%

\subsection{Model C: $Q_{C}=9nH(\protect\rho_{\Lambda}+\protect\rho_{m})$}

For the third model of our analysis, where $Q_{C}=9nH(\rho_{\Lambda}+\rho
_{m})$, the field equations read

\begin{equation}
\frac{d%
\Omega
_{\Lambda}}{d\ln a}=2\Omega_{\Lambda}\left[ 1+\frac{3}{4}(\Omega_{m}+%
\Omega_{\Lambda}\ -1)+\frac{1}{2}\Omega_{m}(1+3w_{m})-\Omega_{\Lambda }%
\right] -3n\Omega_{\Lambda}-3n\Omega_{m},  \label{c1}
\end{equation}

\begin{equation}
\frac{d%
\Omega
_{m}}{d\ln a}=2\Omega_{m}\left( 1+\frac{3}{4}(\Omega_{m}+\Omega_{\Lambda }\
-1)+\frac{1}{2}\Omega_{m}(1+3w_{m})-\Omega_{\Lambda}\right) +3n\Omega
_{\Lambda}+3n\Omega_{m}-3\Omega_{m}(1+w_{m}),  \label{c2}
\end{equation}

The latter dynamical system admits the following critical points 
\begin{equation*}
C_{1}=\{0,0,-1\},~C_{2\pm}=\{\frac{1}{2}\left( 1\pm\sqrt{\frac{x}{(1+w_{m})}}%
\right) ,\frac{1}{2}\left( 1\mp\sqrt{\frac{x}{(1+w_{m})}}\right) ,0\},
\end{equation*}
where we considered $x=1-4n+w_{m}.$

The universe described by the exact solution at the stationary point $C_{1}$
has the same physical quantities with those of points $A_{1}$ and $B_{1}$.
The eigenvalues of the linearized system are%
\begin{align*}
e_{1}\left( C_{1}\right) & =-\frac{1}{2}\left( 2+3w_{m}+3\sqrt {(1+w_{m})x}%
\right) ~ \\
e_{2}\left( C_{1}\right) & =-\frac{1}{2}\left( 2+3w_{m}-3\sqrt {(1+w_{m})x}%
\right)
\end{align*}
from where we can infer that the point is an attractor when $\left\{ \frac {1%
}{12}<n<\frac{1}{2},-\frac{2}{3}<w<-1+4n\right\} $\textbf{,} $\left\{ n>%
\frac{1}{2},-\frac{2}{3},w<1\right\} ,~\left\{ \frac{1}{12}<n<\frac {11}{72}%
,-1+4n<w<\frac{5-36n}{36n-6}\right\} ,~\left\{ \frac{11}{72}<n<\frac{1}{2}%
,-1+4n<w<1\right\} .$

Point $C_{2\pm}~$exists for $\left\{ n<0~\&~w_{m}\leq4n-1\right\} \ $or$\
\left\{ n>0~\&~w_{m}\geqslant4n-1\right\} $ and describes the same physical
solutions with points $A_{3}$ and $B_{3}$. The equation of state parameter
is $w_{m}\left( C_{2\pm}\right) =\frac{1}{2}\left( -1+w_{m}\pm\sqrt{\left(
1+w_{m}\right) x}\right) $. From the linearized system around the critical
points we determine the eigenvalues%
\begin{align*}
e_{1}\left( C_{2\pm}\right) & =\frac{1}{4}\left( 2+3w_{m}\pm 9\sqrt{\left(
1+w_{m}\right) x}\right) \\
& +\frac{1}{4}\sqrt{13-36n\left( 1+w_{m}\right) \mp12\sqrt{\left(
1+w_{m}\right) x}+6w_{m}\left( 5+3w_{m}\mp3\sqrt{\left( 1+w_{m}\right) x}%
\right) },
\end{align*}%
\begin{align*}
e_{2}\left( C_{2\pm}\right) & =\frac{1}{4}\left( 2+3w_{m}\pm 9\sqrt{\left(
1+w_{m}\right) x}\right) \\
& -\frac{1}{4}\sqrt{13-36n\left( 1+w_{m}\right) \mp12\sqrt{\left(
1+w_{m}\right) x}+6w_{m}\left( 5+3w_{m}\mp3\sqrt{\left( 1+w_{m}\right) x}%
\right) }.
\end{align*}

Therefore, point $C_{2-}$ is always unstable while point $C_{2+}$ is
conditionally stable as shown in \ref{tabc02}.

\begin{figure}[ptb]
\centering\includegraphics[width=0.40\textwidth]{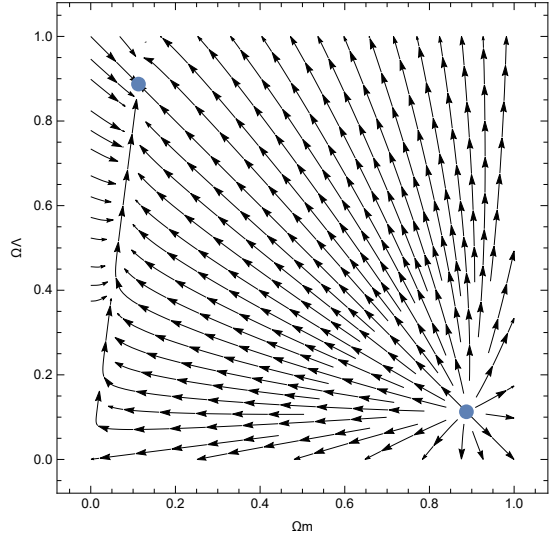}
\caption{Phase space diagram for the dynamical system (\protect\ref{c1}), (%
\protect\ref{c2}). We consider $w_{m}=0$, for $n<1$. The unique attractor is
point $C_{2}$. }
\label{fig:fig9}
\end{figure}

\begin{figure}[ptb]
\centering\includegraphics[width=0.7\textwidth]{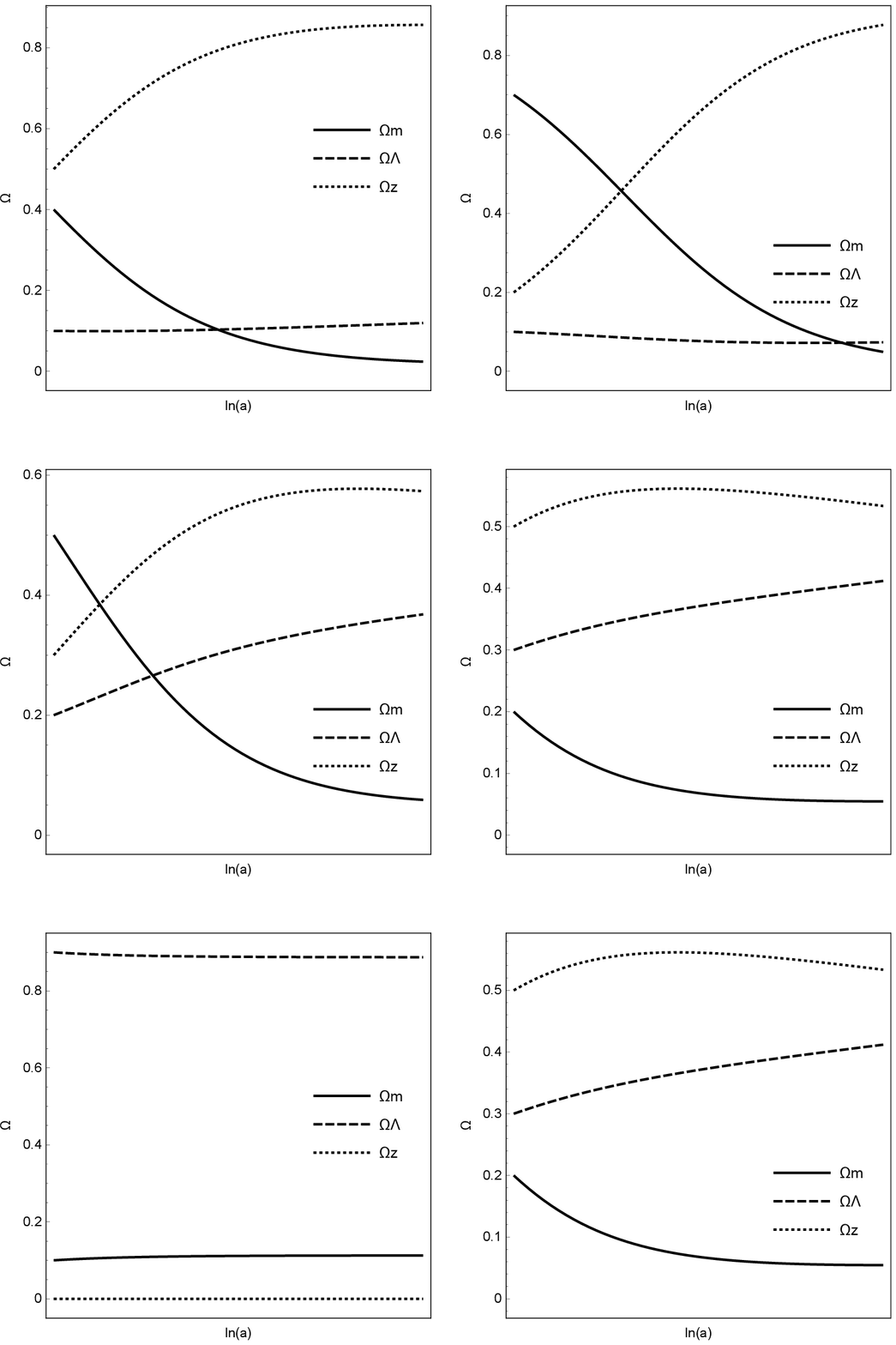}
\caption{Evolution diagrams with time, for the various densities for the
dynamical system (\protect\ref{c1}), (\protect\ref{c2}). We consider the
initial conditions (a) $\Omega _{m}=0.4,$ $\Omega_{\Lambda}=0.1$ (b) $%
\Omega_{m}=0.7$, $\Omega_{\Lambda}=0.1$ (c) $\Omega_{m}=0.5$, $%
\Omega_{\Lambda}=0.2$ (d) $\Omega_{m}=0.2$, $\Omega_{\Lambda}=0.3$ (e) $%
\Omega_{m}=0.1$, $\Omega_{\Lambda}=0.9$ (f) $\Omega_{m}=0.2$, $%
\Omega_{\Lambda}=0.3$, for $n<1$ and $w_{m}=0$.}
\label{fig:fig11}
\end{figure}

The above results are summarized in Tables \ref{tabc01},\ref{tabc02} and \ref%
{accelC2}. Moreover, in Figs. \ref{fig:fig9}, \ref{fig:fig11}~the evolution
of trajectories for the dynamical system our study in phase space are
presented.

\begin{table}[tbp] \centering%
\caption{Stationary points and physical parameters for the interaction model
C. }%
\begin{tabular}{ccccc}
\hline\hline
\textbf{Point} & $\left( \mathbf{%
\Omega
}_{m}\mathbf{,%
\Omega
}_{\Lambda}\mathbf{,%
\Omega
}_{z}\right) $ & \textbf{Existence} & $\mathbf{w}_{tot}$ & \textbf{%
Acceleration} \\ \hline
$C_{1}$ & $\left( 0,0,-1\right) $ & Always & $-\frac{5}{6}$ & Yes \\ 
$C_{2\pm}$ & $\left( \frac{1}{2}\left( 1\pm\sqrt{\frac{x}{(1+w_{m})}}\right)
,\frac{1}{2}\left( 1\mp\sqrt{\frac{x}{(1+w_{m})}}\right) ,0\right) $ & $%
w_{m}\neq0,n<0\ \&\ w_{m}\leq4n-1$ & $\frac{1}{2}\left( w_{m}-1\pm\sqrt{%
\left( 1+w_{m}\right) x}\right) $ & see\ \ref{accelC2} \\ 
&  & $w_{m}\neq0,n>0\ \&\ w_{m}\geqslant4n-1$ &  &  \\ \hline
\end{tabular}
\label{tabc01}%
\end{table}%

\begin{table}[tbp] \centering%
\caption{Stationary points and stability conditions for the interaction
model C. }%
\begin{tabular}{cc}
\hline\hline
\textbf{Point} & \textbf{Stability} \\ \hline
$C_{1}$ & $\left\{ \frac{1}{12}<n<\frac{1}{2},-\frac{2}{3}<w<-1+4n\right\} $%
, $\left\{ n>\frac{1}{2},-\frac{2}{3},w<1\right\} $ \\ \hline
& $~\left\{ \frac{1}{12}<n<\frac{11}{72},-1+4n<w<\frac{5-36n}{36n-6}\right\} 
$ \\ \hline
& $\left\{ \frac{11}{72}<n<\frac{1}{2},-1+4n<w<1\right\} $ \\ \hline
$C_{2-}$ & Unstable \\ \hline
$C_{2+}$ & $n<0:~w_{m}<-1+4n,~-1<w_{m}<\frac{5-36n}{36n-6},~\frac {5-36n}{%
36n-6}<w_{m}<-\frac{2}{3}~$ \\ 
& $0<n<\frac{1}{12}:w_{m}<-1,-1+4n<w_{m}<\frac{5-36n}{36n-6},\frac {5-36n}{%
36n-6}<w_{m}<-\frac{2}{3}~$ \\ 
& $\frac{1}{12}<n<\frac{1}{6}:w_{m}<-1$ \\ 
& \thinspace$n>\frac{1}{6}:w_{m}<\frac{5-36n}{36n-6},\frac{5-36n}{36n-6}%
<w_{m}<-1$ \\ 
& \thinspace$n<\frac{1}{12}:w_{m}=\frac{5-36n}{36n-6}$ \\ 
& $-\frac{2}{3}\leq w_{m}<1,~$\thinspace$n<\frac{6w_{m}+5}{36\left(
1+w_{m}\right) }$. \\ \hline
\end{tabular}
\label{tabc02}%
\end{table}%

\begin{table}[tbp] \centering%
\caption{Acceleration conditions for the interaction model C for point
$C_{2±}$. }%
\begin{tabular}{cc}
\hline\hline
\textbf{Point} & \textbf{Acceleration} \\ \hline
$C_{2\pm}$ & $n=0$ \\ 
& $n>0$ \& $w_{m}\leq-1$ or $n<\frac{1}{3}$ and $4n-1\leq w_{m}$ \\ 
& $\frac{2}{3}>n\geqslant\frac{1}{3}$ and $w_{m}>\frac{4}{9n-6}-1$ \\ 
& $n<0~$and $[4n-1\geqslant w_{m}~$or $w_{m}\geqslant-1]$ \\ \hline
\end{tabular}
\label{accelC2}%
\end{table}%

\subsection{Model D: $Q_{D}=$ $9nH^{3}\Omega_{z}$}

In this scenario we shall consider an interaction of the form, $Q=Q\left(
\Omega_{z}\right) $, that of course due to the constraint equation (\ref{con}%
) means that%
\begin{equation*}
Q=Q\left( \Omega_{m},\Omega_{\Lambda}\right)
\end{equation*}
So, if we consider the interaction term to be $Q=$ $9nH^{3}\Omega_{z}$ then
it follows%
\begin{equation*}
Q=9nH^{3}\left( \Omega_{m}+\Omega_{\Lambda}-1\right) .
\end{equation*}

and our system is now

\begin{equation}
\frac{d%
\Omega
_{\Lambda}}{d\ln a}=2\Omega_{\Lambda}\left[ 1+\frac{3}{4}(\Omega_{m}+%
\Omega_{\Lambda}\ -1)+\frac{1}{2}\Omega_{m}(1+3w_{m})-\Omega_{\Lambda }%
\right] -3n(\Omega_{m}+\Omega_{\Lambda}-1),  \label{d1}
\end{equation}

\begin{equation}
\frac{d%
\Omega
_{m}}{d\ln a}=2\Omega_{m}\left[ 1+\frac{3}{4}(\Omega_{m}+\Omega_{\Lambda }\
-1)+\frac{1}{2}\Omega_{m}(1+3w_{m})-\Omega_{\Lambda}\right] +3n(\Omega
_{m}+\Omega_{\Lambda}-1)-3\Omega_{m}(1+w_{m}),  \label{d2}
\end{equation}

The dynamical system (\ref{d1}), (\ref{d2}) admits three critical points
with coordinates 
\begin{equation*}
D_{1}=\{0,1,0\},D_{2}=\{1,0,0\},D_{4}=\left\{ 6n\alpha,6n\alpha
(5+6w_{m}),(5+6w_{m})\alpha\right\}
\end{equation*}
where$~\alpha=\left( 36n(1+w_{m})-(5+6w_{m})\right) ^{-1}.$ Point $D_{1}$
describes a de Sitter universe with equation of state parameter~$%
w_{tot}\left( D_{1}\right) =-1$, where only the varying vacuum\ term
contributes in the evolution of the universe. The eigenvalues are derived to
be $\{-\frac {1}{2},-\ 3(1+w_{m})\},$ so for $w_{m}\geq-1$ the point is
always an attractor and this point is of great physical interest.

Point $D_{2}~$describes a universe dominated by matter, $w_{tot}\left(
D_{2}\right) =w_{m}$, and the exact solution at the point corresponds to an
accelerated universe when $w_{m}\leq-\frac{1}{3}$. The eigenvalues of the
linearized system are $\left\{ 3(1+w_{m}),\frac{(5+6w_{m})}{2}\right\} ,$
from where we observe that this point is an attractor only when $w_{m}<-1$.

Point $D_{3}$ exists when $~n<0,w_{m}>-\frac{5}{6}$ or $n>\frac{1}{6}%
,0<w_{m}+\frac{5}{6}<n$ and it corresponds to a universe of two fluids and
the contribution of the geometrical background of Finsler Randers that is
always accelerating, that is, $w_{tot}\left( D_{3}\right) =-\frac{5}{6}$.
Given that we consider the values of $n$ very small this solution describes
a universe where matter decays in vacuum. The eigenvalues of the linearized
system near the stationary point are $\left\{ \frac{1}{2},-\frac{(5+6w_{m})}{%
2}\right\} $, so point $D_{3}$ is always a source, since one of the
eigenvalues has always positive real part.

\begin{figure}[ptb]
\centering\includegraphics[width=0.40\textwidth]{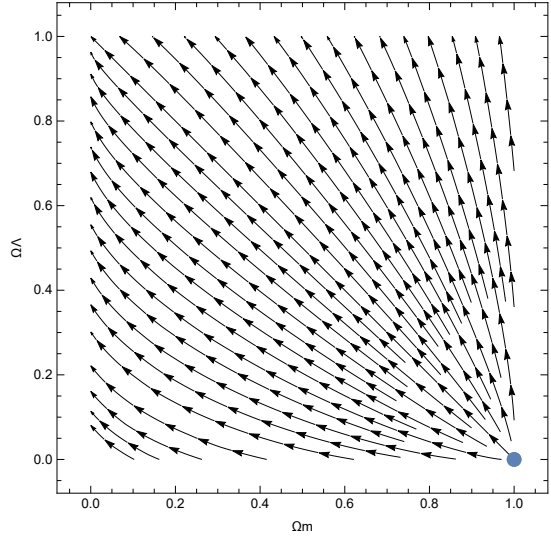}
\caption{Phase space diagram for the dynamical system (\protect\ref{d1}), (%
\protect\ref{d2}). We consider $\Omega_{m}=0,\;w_{m}=0$, for $n<1$. The
unique attractor is the point $D_{1}$. }
\label{fig:fig13}
\end{figure}

\begin{figure}[ptb]
\centering\includegraphics[width=0.7\textwidth]{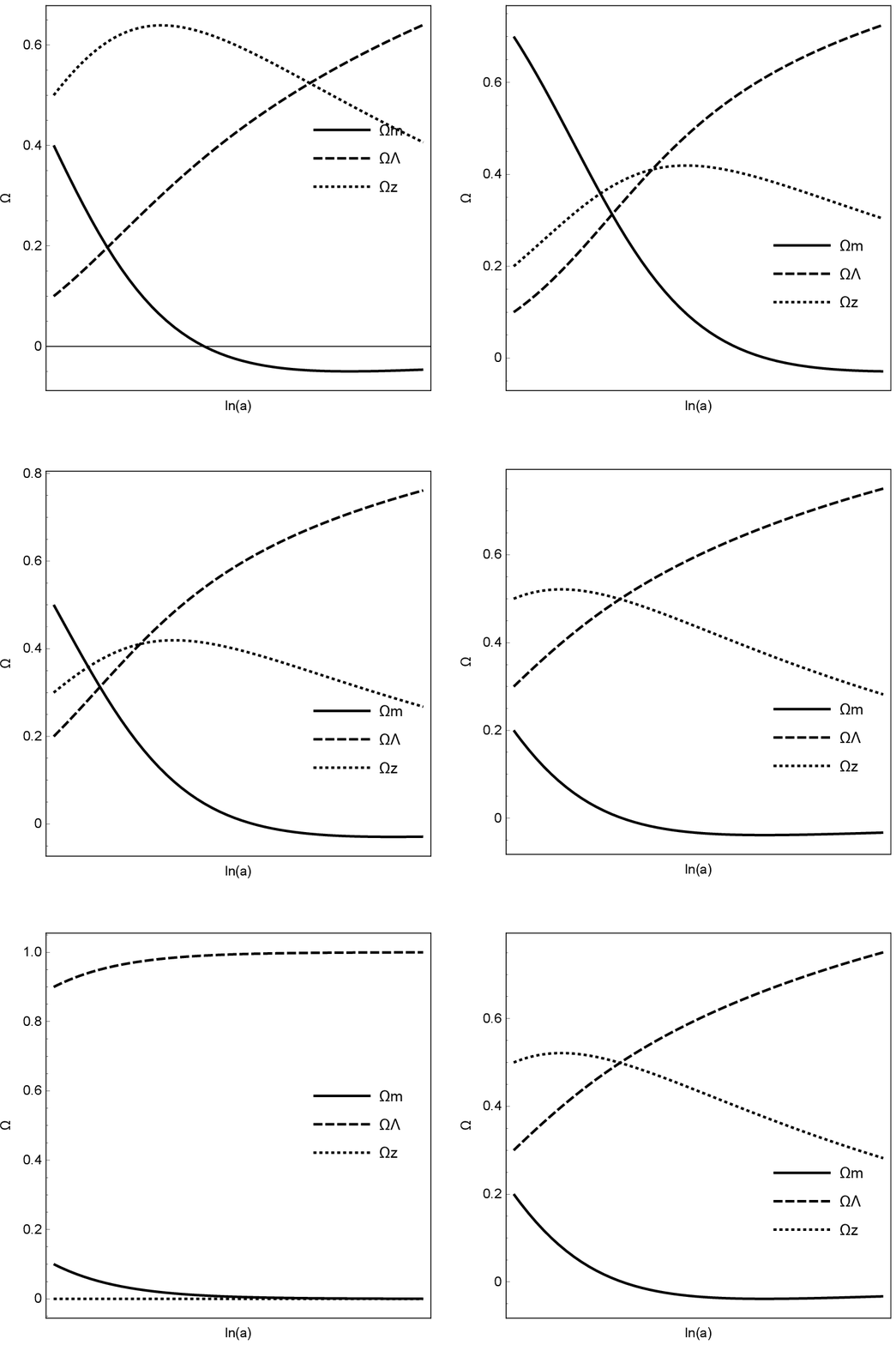}
\caption{Evolution diagrams with time, for various energy densities of the
dynamical system (\protect\ref{d1}), (\protect\ref{d2}). We consider the
initial conditions (a) $\Omega _{m}=0.4,$ $\Omega_{\Lambda}=0.1$ (b) $%
\Omega_{m}=0.7$, $\Omega_{\Lambda}=0.1$ (c) $\Omega_{m}=0.5$, $%
\Omega_{\Lambda}=0.2$ (d) $\Omega_{m}=0.2$, $\Omega_{\Lambda}=0.3$ (e) $%
\Omega_{m}=0.1$, $\Omega_{\Lambda}=0.9$ (f) $\Omega_{m}=0.2$, $%
\Omega_{\Lambda}=0.3$, for $n<1$ and $w_{m}=0$.}
\label{fig:fig15}
\end{figure}

The above results are summarized in Tables \ref{tabd01} and \ref{tabd02}. In
Figs. \ref{fig:fig13},~\ref{fig:fig15}~the evolution of real trajectories
for the dynamical system our study in phase space are presented.

\begin{table}[tbp] \centering%
\caption{Stationary points and physical parameters for the interaction model
D. }%
\begin{tabular}{ccccc}
\hline\hline
\textbf{Point} & $\left( \mathbf{%
\Omega
}_{m}\mathbf{,%
\Omega
}_{\Lambda}\mathbf{,%
\Omega
}_{z}\right) $ & \textbf{Existence} & $\mathbf{w}_{tot}$ & \textbf{%
Acceleration} \\ \hline
$D_{1}$ & $\left( 0,1,0\right) $ & Always & $-1$ & Yes \\ 
$D_{2}$ & $\left( 1,0,0\right) $ & Always & $w_{m}$ & $w_{m}\leq-\frac{1}{3}$
\\ 
$D_{3}$ & $\left( 6n\alpha,6n\alpha(5+6w_{m}),(5+6w_{m})\alpha\right) $ & $%
n<0,w_{m}>-\frac{5}{6}$ or $n>\frac{1}{6},0<w_{m}+\frac{5}{6}<n$ & $-\frac {5%
}{6}$ & Yes \\ \hline\hline
\end{tabular}
\label{tabd01}%
\end{table}%

\begin{table}[tbp] \centering%
\caption{Stationary points and stability conditions for the interaction
model D. }%
\begin{tabular}{ccc}
\hline\hline
\textbf{Point} & \textbf{Ei}$\text{\textbf{genvalues}}$ & \textbf{Stability}
\\ \hline\hline
$D_{1}$ & $\{-\frac{1}{2},-\ 3(1+w_{m})\}$ & $w_{m}>-1$ \\ 
$D_{2}$ & $\left\{ 3(1+w_{m}),\frac{(5+6w_{m})}{2}\right\} $ & $w_{m}<-1$ \\ 
$D_{3}$ & $\left\{ \frac{1}{2},-\frac{(5+6w_{m})}{2}\right\} $ & Source \\ 
\hline\hline
\end{tabular}
\label{tabd02}%
\end{table}%

\subsection{Model E: $Q_{E}=$ $9nH^{3}\Omega_{z}+9mH\protect\rho_{m}$}

Our system is now

\begin{equation}
\frac{d%
\Omega
_{\Lambda}}{d\ln a}=2\Omega_{\Lambda}\left[ 1+\frac{3}{4}(\Omega_{m}+%
\Omega_{\Lambda}\ -1)+\frac{1}{2}\Omega_{m}(1+3w_{m})-\Omega_{\Lambda }%
\right] -3n(\Omega_{m}+\Omega_{\Lambda}-1)-3m\Omega_{m},  \label{e11}
\end{equation}%
\begin{align}
\frac{d%
\Omega
_{m}}{d\ln a} & =2\Omega_{m}\left[ 1+\frac{3}{4}(\Omega_{m}+\Omega
_{\Lambda}\ -1)+\frac{1}{2}\Omega_{m}(1+3w_{m})-\Omega_{\Lambda}\right]
\label{e22} \\
& +3n(\Omega_{m}+\Omega_{\Lambda}-1)+3m\Omega_{m}-3\Omega_{m}(1+w_{m}), 
\notag
\end{align}

The dynamical system (\ref{e11}), (\ref{e22}), admits three critical points
with coordinates 
\begin{equation*}
E_{1}=\{0,1,0\},~E_{2}=\{1,0,0\},E_{3}=\{6nb,6nb(5+6w_{m}),(5+6w_{m}-6b)b\}.
\end{equation*}
where $b=\left( 36n(1+w_{m})-(5+6w_{m})+6m\right) ^{-1}.$ Point $E_{1}$
describes a de Sitter universe with equation of state parameter~$%
w_{tot}\left( E_{1}\right) =-1$, where only the varying vacuum\ term
contributes in the evolution of the universe. The eigenvalues are derived to
be $\{-\frac {1}{2},-\ 3(1+w_{m}-m)\},$ so for $w_{m}>m-1$ the point is
always an attractor and thus this solution is of great physical interest.

Point\textbf{\ }$E_{2}~$describes a universe dominated by the varying vacuum
and matter; when $w_{m}=0$, point $E_{2}$ describes the $\Lambda$-CDM
universe in the FR theory. The equation of state parameter is derived $%
w_{tot}\left( E_{2}\right) =w_{m}-m$, so this point describes an accelerated
universe when $w_{m}\leq m-\frac{1}{3}$. The eigenvalues of the linearized
system are\ $\left\{ 3(1+w_{m}-m),\frac{(5+6w_{m})-6m}{2}\right\} $ from
where we can infer that the point is stable for $w_{m}<m-1.$Given though the
existence condition $m-1\leq$ $w_{m}$ we consider the point to be unstable.

Point $E_{3}$ exists when $n,$ $m$ and~$w_{m}$ are constrained as presented
in Table \ref{tabe03e}. Similarly with point $D_{\mathbf{3}}$ this point
corresponds to a universe of two fluids and the contribution of the
geometrical background of Finsler Randers that is always accelerating$%
(w_{tot}\left( E_{3}\right) =-\frac{5}{6})$. The eigenvalues of the
linearized system near the stationary point are $\left\{ \frac{1}{2},-\frac{%
(5+6w_{m})}{2}+3m\right\} $, so point $E_{3}$ is\ always a source.

\begin{figure}[ptb]
\centering\includegraphics[width=0.40\textwidth]{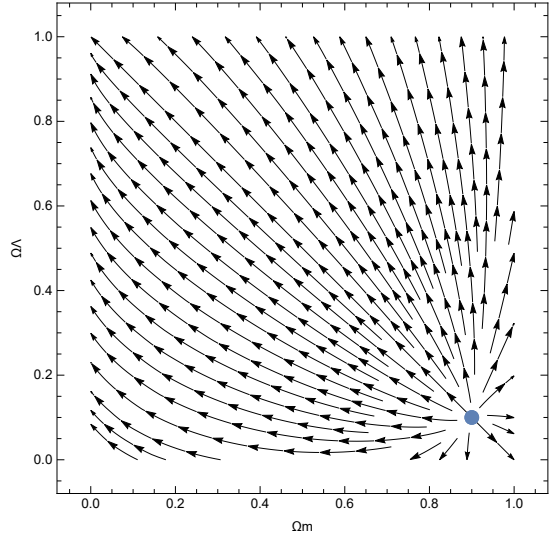}
\caption{Phase space diagram for the dynamical system (\protect\ref{e11}), (%
\protect\ref{e22}). We consider $w_{m}=0$, for $n<1$. The unique attractor
is point $E_{1}$. }
\label{fig:fig17}
\end{figure}

\begin{figure}[ptb]
\centering\includegraphics[width=0.7\textwidth]{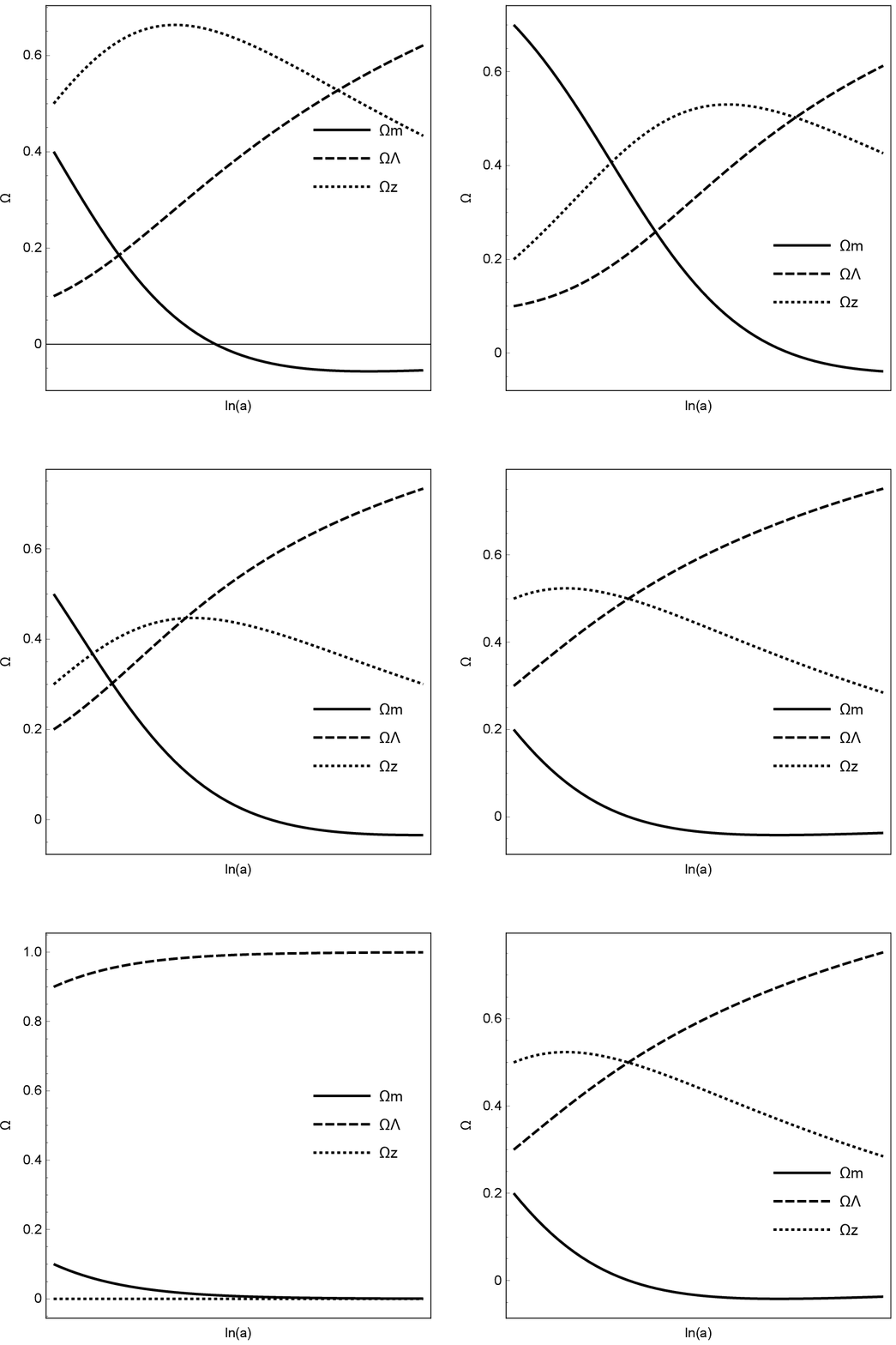}
\caption{Evolution diagrams with time, for various energy densities of the
dynamical system (\protect\ref{e11}), (\protect\ref{e22}). We consider the
initial conditions (a) $\Omega _{m}=0.4,$ $\Omega_{\Lambda}=0.1$ (b) $%
\Omega_{m}=0.7$, $\Omega_{\Lambda}=0.1$ (c) $\Omega_{m}=0.5$, $%
\Omega_{\Lambda}=0.2$ (d) $\Omega_{m}=0.2$, $\Omega_{\Lambda}=0.3$ (e) $%
\Omega_{m}=0.1$, $\Omega_{\Lambda}=0.9$ (f) $\Omega_{m}=0.2$, $%
\Omega_{\Lambda}=0.3$, for $n<1$ and $w_{m}=0$.}
\label{fig:fig19}
\end{figure}

The above results are summarized in Tables \ref{tabe01e} and \ref{tabe02e}.
The trajectories of the dynamical system in the phase space are presented in
Figs. \ref{fig:fig17}, \ref{fig:fig19}.

\begin{table}[tbp] \centering%
\caption{Stationary points and physical parameters for the interaction model
E. }%
\begin{tabular}{ccccc}
\hline\hline
\textbf{Point} & $\left( \mathbf{%
\Omega
}_{m}\mathbf{,%
\Omega
}_{\Lambda}\mathbf{,%
\Omega
}_{z}\right) $ & \textbf{Existence} & $\mathbf{w}_{tot}$ & \textbf{%
Acceleration} \\ \hline
$E_{1}$ & $\left( 0,1,0\right) $ & Always & $-1$ & Yes \\ 
$E_{2}$ & $\left( 1-\frac{m}{1+w_{m}},\frac{m}{1+w_{m}},0\right) $ & $%
w_{m}>-1,0\leq m\leq$ $1+w_{m}$ & $w_{m}-m$ & $w_{m}\leq m-\frac{1}{3}$ \\ 
$E_{3}$ & $\left( 6nb,6nb(5+6w_{m}),(5+6w_{m}-6m)b\right) $ & See Table \ref%
{tabe03e} & $-\frac{5}{6}$ & Yes \\ \hline\hline
\end{tabular}
\label{tabe01e}%
\end{table}%

\begin{table}[tbp] \centering%
\caption{Stationary points and stability conditions for the interaction
model E. }%
\begin{tabular}{ccc}
\hline\hline
\textbf{Point} & \textbf{Ei}$\text{\textbf{genvalues}}$ & \textbf{Stability}
\\ \hline\hline
$E_{1}$ & $\{-\frac{1}{2},-\ 3(1+w_{m}-m)\}$ & $w_{m}>m-1$ \\ 
$E_{2}$ & $\left\{ 3(1+w_{m}-m),\frac{(5+6w_{m})-6m}{2}\right\} $ & $%
w_{m}<-1+m$ \\ 
$E_{3}$ & $\left\{ \frac{1}{2},-\frac{(5+6w_{m})}{2}+3m\right\} $ & Source
\\ \hline\hline
\end{tabular}
\label{tabe02e}%
\end{table}%
\begin{table}[tbp] \centering%
\caption{Existence conditions for the stationary point $E_{4}$}%
\begin{tabular}{ccc}
\hline\hline
\textbf{Point} & \textbf{Existence} & \textbf{Existence} \\ \hline
$E_{3}$ & $m<0$ & $n<0$ and $w_{m}\geq-\frac{5}{6}$ \\ 
&  & $m+n=\frac{1}{6}$ and $5+6w_{m}=\frac{6m}{6n-1}$ \\ 
&  & $n=0$ and $(w_{m}<m-\frac{5}{6}$ \ or $w_{m}>m-\frac{5}{6})$ \\ 
&  & $m+n>\frac{1}{6}$and $6m+6n\geq5+6w_{m}\geq\frac{6m}{6n-1}$ \\ 
\cline{2-3}
& $0<m\leq\frac{1}{6}$ & $5+6w_{m}>0$ and [($n>0,m+n\leq\frac{1}{6}%
,5+6w_{m}\leq\frac{6m}{6n-1}$) or ($m+n>\frac{1}{6},6m+6n\leq5+6w_{m}$)] \\ 
\cline{2-3}
& $m>\frac{1}{6}$ & $m+n\leq\frac{1}{6}\ $and $5+6w_{m}\geq\frac{6m}{6n-1}$
\\ 
&  & $m+n\leq\frac{1}{6}$and $n<0,6m+6n\leq5+6w_{m}$ \\ 
&  & $n>0$ and $0<5+6w_{m}\leq6m+6n$ \\ \cline{2-3}
& $m=0$, $5+6w_{m}>0$ & $n<0~$or $[n\geq\frac{1}{6}~$and $n\geq\frac{5}{6}%
+w_{m}]$ \\ \hline\hline
\end{tabular}
\label{tabe03e}%
\end{table}%

\subsection{Model F: $Q_{F}=$ $9nH^{3}\Omega_{z}+9mH\protect\rho_{\Lambda}$}

Our system is now

\begin{equation}
\frac{d%
\Omega
_{\Lambda}}{d\ln a}=2\Omega_{\Lambda}\left[ 1+\frac{3}{4}(\Omega_{m}+%
\Omega_{\Lambda}\ -1)+\frac{1}{2}\Omega_{m}(1+3w_{m})-\Omega_{\Lambda }%
\right] -3n(\Omega_{m}+\Omega_{\Lambda}-1)-3m\Omega_{\Lambda},  \label{f1}
\end{equation}%
\begin{align}
\frac{d%
\Omega
_{m}}{d\ln a} & =2\Omega_{m}\left[ 1+\frac{3}{4}(\Omega_{m}+\Omega
_{\Lambda}\ -1)+\frac{1}{2}\Omega_{m}(1+3w_{m})-\Omega_{\Lambda}\right]
\label{f2} \\
& +3n(\Omega_{m}+\Omega_{\Lambda}-1)+3m\Omega_{\Lambda}-3\Omega_{m}(1+w_{m}),
\notag
\end{align}

The dynamical system (\ref{f1}), (\ref{f2}) admits three critical points
with coordinates 
\begin{equation*}
F_{1}=\{0,1,0\},F_{2}=\{1,0,0\},F_{3}=\left\{
6nc,6nc(5+6w_{m}),(5+6w_{m})(1-6m)c\right\} ,
\end{equation*}
where $c=\left( 36n(1+w_{m})+(5+6w_{m})(6m-1)\right) ^{-1}.$ Point $F_{1}~$%
describes a universe dominated by matter, $w_{tot}\left( F_{1}\right) =w_{m}$%
, and the exact solution at the point corresponds to an accelerated universe
for $w_{m}\leq-\frac{1}{3}$. The eigenvalues of the linearized system are $\{%
\frac{(5+6w_{m})}{2},\ 3(1+w_{m}-m)\},$ from where we observe that this
point is an attractor only when $w_{m}<-\frac{5}{6}\ $and $1+w_{m}<m$. Thus
this point provides a viable scenario of a matter dominated universe.

Point\textbf{\ }$F_{2}~$describes a universe dominated by the varying vacuum
and matter; when $w_{m}=0$, point $F_{3}$ describes the $\Lambda$-CDM
universe in the FR theory. The equation of state parameter is derived $%
w_{tot}\left( F_{2}\right) =m-1$, so this point describes an accelerated
universe when $m\leq\frac{2}{3}$. The eigenvalues of the linearized system
are\ $\left\{ -3(1+w_{m}-m),3m-\frac{1}{2}\right\} $ from where we can infer
that the point is stable for $m<\frac{1}{6}$ and $1+w_{m}>m.$We observe that
for the theoretical values of m (very small) this is a stable point that
describes an accelerated universe and thus it is extremely interesting from
a physical point of view.

\begin{figure}[ptb]
\centering\includegraphics[width=0.40\textwidth]{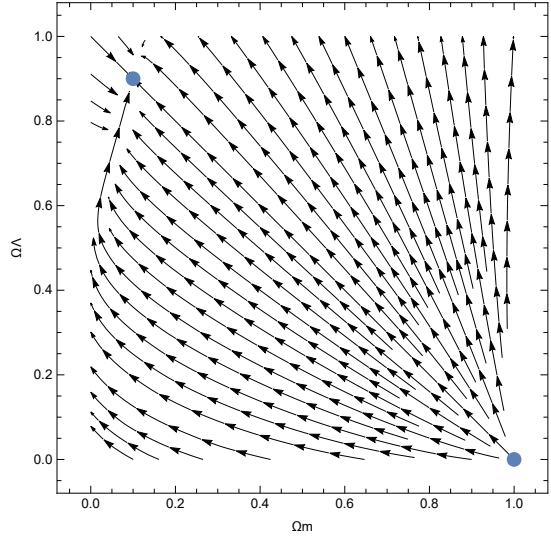}
\caption{Phase space diagram for the dynamical system (\protect\ref{f1}), (%
\protect\ref{f2}). We consider $w_{m}=0$, for $n<1$. The unique attractor is
point $F_{3}$. }
\label{fig:fig21}
\end{figure}

\begin{figure}[ptb]
\centering\includegraphics[width=0.7\textwidth]{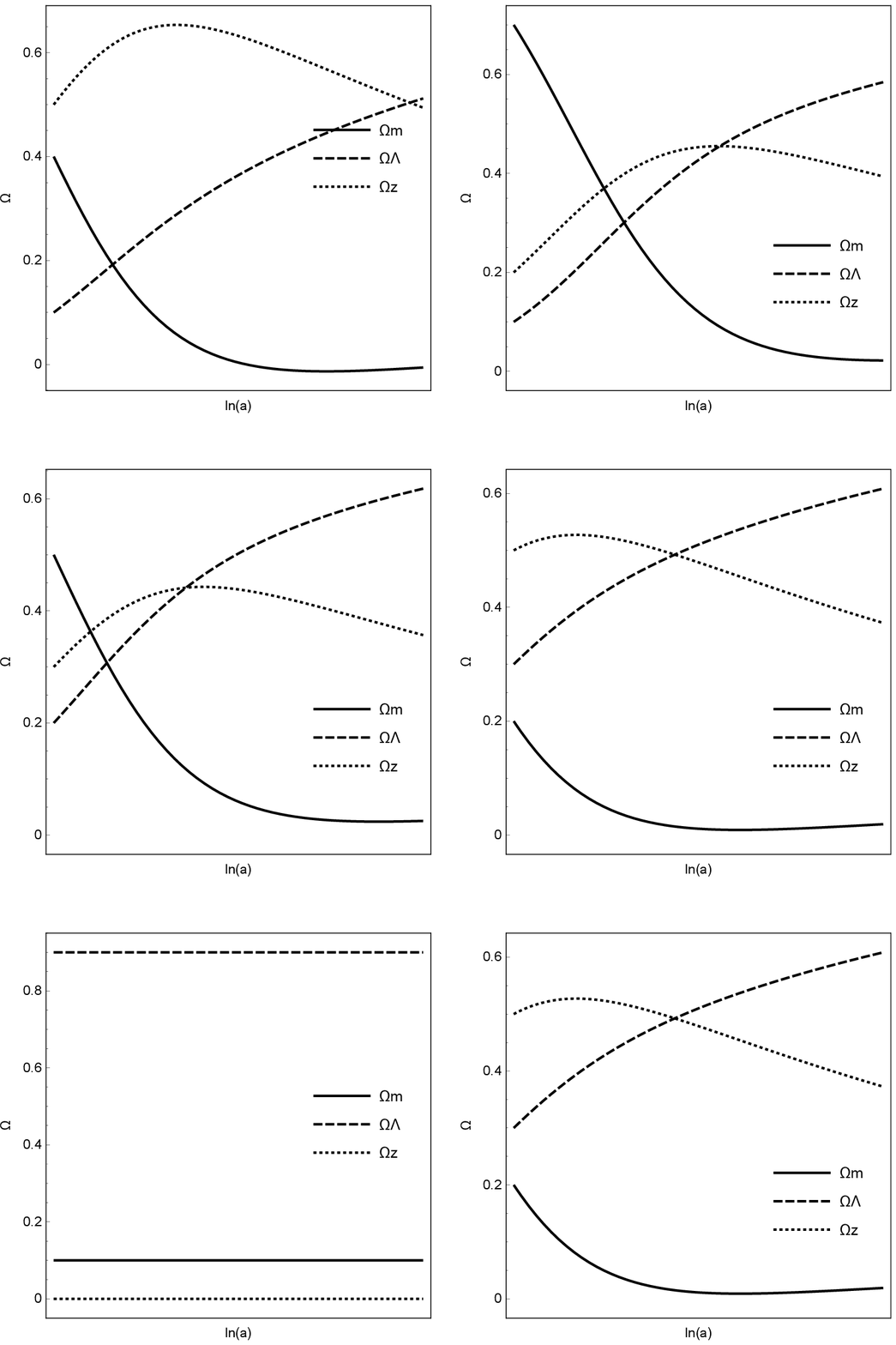}
\caption{Evolution diagrams with time, for various energy densities of the
dynamical system (\protect\ref{f1}), (\protect\ref{f2}). We consider the
initial conditions (a) $\Omega _{m}=0.4,$ $\Omega_{\Lambda}=0.1$ (b) $%
\Omega_{m}=0.7$, $\Omega_{\Lambda}=0.1$ (c) $\Omega_{m}=0.5$, $%
\Omega_{\Lambda}=0.2$ (d) $\Omega_{m}=0.2$, $\Omega_{\Lambda}=0.3$ (e) $%
\Omega_{m}=0.1$, $\Omega_{\Lambda}=0.9$ (f) $\Omega_{m}=0.2$, $%
\Omega_{\Lambda}=0.3$, for $n<1$ and $w_{m}=0$.}
\label{fig:fig23}
\end{figure}

The existence conditions of point $F_{3}$ are given in Table \ref{tabe03f}.
Similar with point $E_{\mathbf{3}},$ it corresponds to a universe of two
fluids and the contribution of the geometrical background of Finsler Randers
that is always accelerating$,~$that is, $w_{tot}\left( F_{3}\right) =-\frac{5%
}{6}$. The eigenvalues of the linearized system near the stationary point
are $\left\{ \frac{1}{2}-3m,-\frac{(5+6w_{m})}{2}\right\} $, so point $F_{3}$
is an attractor for $m>\frac{1}{6}~$and $w_{m}>-\frac{5}{6}$. The above
results are summarized in Tables \ref{tabe01f} and \ref{tabe02f}. In Figs. %
\ref{fig:fig21}, \ref{fig:fig23}~the evolution of trajectories for the
dynamical system our study in phase space are presented.

\begin{table}[tbp] \centering%
\caption{Stationary points and physical parameters for the interaction model
F. }%
\begin{tabular}{ccccc}
\hline\hline
\textbf{Point} & $\left( \mathbf{%
\Omega
}_{m}\mathbf{,%
\Omega
}_{\Lambda}\mathbf{,%
\Omega
}_{z}\right) $ & \textbf{Existence} & $\mathbf{w}_{tot}$ & \textbf{%
Acceleration} \\ \hline
$F_{1}$ & $\left( 1,0,0\right) $ & Always & $w_{m}$ & $w_{m}\leq-\frac{1}{3}$
\\ 
$F_{2}$ & $\left( \frac{m}{1+w_{m}},1-\frac{m}{1+w_{m}},0\right) $ & $%
w_{m}>-1,0\leq m\leq$ $1+w_{m}$ & $-1+m$ & $m\leq\frac{2}{3}$ \\ 
$F_{3}$ & $\left( 6nc,6nc(5+6w_{m}),(5+6w_{m})(6m-1)c\right) $ & See Table %
\ref{tabe03f} & $-\frac{5}{6}$ & Yes \\ \hline\hline
\end{tabular}
\label{tabe01f}%
\end{table}%

\begin{table}[tbp] \centering%
\caption{Stationary points and stability conditions for the interaction
model F. }%
\begin{tabular}{ccc}
\hline\hline
\textbf{Point} & \textbf{Ei}$\text{\textbf{genvalues}}$ & \textbf{Stability}
\\ \hline\hline
$F_{1}$ & $\{\frac{(5+6w_{m})}{2},\ 3(1+w_{m}-m)\}$ & $w_{m}<-\frac{5}{6}\ $%
and $1+w_{m}<m$ \\ 
$F_{2}$ & $\left\{ -3(1+w_{m}-m),3m-\frac{1}{2}\right\} $ & Attractor for$~m<%
\frac{1}{6}$ and $1+w_{m}>m$ \\ 
$F_{3}$ & $\left\{ \frac{1}{2}-3m,-\frac{(5+6w_{m})}{2}\right\} $ & 
Attractor for$~m>\frac{1}{6}~$and $w_{m}>-\frac{5}{6}$ \\ \hline\hline
\end{tabular}
\label{tabe02f}%
\end{table}%

\begin{table}[tbp] \centering%
\caption{Existence conditions for the stationary point $F_{4}$}%
\begin{tabular}{ccc}
\hline\hline
\textbf{Point} & \textbf{Existence} & \textbf{Existence} \\ \hline
$F_{3}$ & $5+6w_{m}\geq0$ & $m=\frac{1}{6},n\neq0$ \\ 
&  & for $m<\frac{1}{6},~n<0$ or $m+n>\frac{1}{6}$ \\ 
&  & for $m>\frac{1}{6},~n>0$ or $m+n<\frac{1}{6}$ \\ 
& $5+6w_{m}\leq0$ & $n>0$ or $m+n<\frac{1}{6}$ \\ 
&  & $n<0$ or $m+n>\frac{1}{6}$ \\ 
& $m\neq\frac{1}{6}$ & $m+n=\frac{1}{6}$~or $n=0$, $5+6w_{m}\neq 0$ \\ 
\hline\hline
\end{tabular}
\label{tabe03f}%
\end{table}%

\subsection{Model G: $Q_{G}=-3\left( 1+\frac{3}{2}w_{m}\right) \Omega
_{z}\Omega_{m}H^{3}$.}

For the last model that we consider the term that is intrinsically by the FR
model, namely $Q_{G}=-3\left( 1+\frac{3}{2}w_{m}\right) \Omega_{z}\Omega
_{m}H^{3}$, and the field equations are expressed as follows.

\begin{equation}
\frac{d%
\Omega
_{\Lambda}}{d\ln a}=\ \frac{1}{2}[\Omega_{\Lambda}-\Omega_{\Lambda}^{2}+%
\Omega_{m}\ (2+3w_{m})(\Omega_{m}-1)+\Omega_{\Lambda}\Omega_{m}(7+9w_{m})]
\label{g1}
\end{equation}

\begin{equation}
\frac{d%
\Omega
_{m}}{d\ln a}=-\frac{3}{2}\Omega_{m}(1+3w_{m})(1+\Omega_{\Lambda}-\Omega_{m})
\label{g2}
\end{equation}

The dynamical system (\ref{g1}),(\ref{g2}) admits four critical points with
coordinates $\{%
\Omega
_{m},%
\Omega
_{\Lambda},%
\Omega
_{z}\}$%
\begin{equation*}
G_{1}=\{0,0,-1\},~G_{2}=\{0,1,0\},~G_{3}=\{1,0,0\},~G_{4}=\{-\frac{1}{%
4+6w_{m}},-\frac{5+6w_{m}}{4+6w_{m}},-2-\frac{2}{4+6w_{m}}\}
\end{equation*}

Point $G_{1}$ always exists and describes an empty universe with equation of
state parameter~$w_{tot}\left( G_{1}\right) =-\frac{5}{6}.$ The universe
accelerates with the contribution of the extra term introduced due to the
Finsler-Randers Geometry. The eigenvalues of the linearized system near to
point $G_{1}$ are $\{\frac{1}{2},-\frac{3}{2}(1+w_{m})\}$, and thus the
point is always a source.

Point $G_{2}~$describes a de Sitter universe with equation of state
parameter~$w_{tot}\left( G_{2}\right) =-1$, where only the $\Lambda$\ term
contributes in the evolution of the universe. The eigenvalues are derived to
be $\{-\frac{1}{2},-\frac{3}{2}(1+w_{m})\},$ from where we can infer that
the point is an attractor when $w_{m}>-1$. Thus this point is of great
physical interest.

Point $G_{3}$ always exists and describes a matter dominated universe that
is accelerating for $w_{m}\leq-\frac{1}{3}.$The eigenvalues of the
linearized system are\ $\{3(1+w_{m}),\frac{(5+6w_{m})}{2}\}$ and thus can be
stable only for $w_{m}<-1.$

Point $G_{4}$ exists only for $w_{m}=-\frac{5}{6}$ in which case it again
describes a matter dominated universe, but in this case it is always
accelerating. By studying its eigenvalues for $w_{m}=-\frac{5}{6}$ though we
deduce that the point is always unstable.

The above results are summarized in Tables \ref{tab01g} and \ref{tab01g}. In
addition in the Figs. \ref{fig:fig1},\ref{fig:fig3} the evolution of
trajectories for the dynamical system our study in phase space are presented.

\begin{figure}[ptb]
\centering
\includegraphics[width=0.4\textwidth]{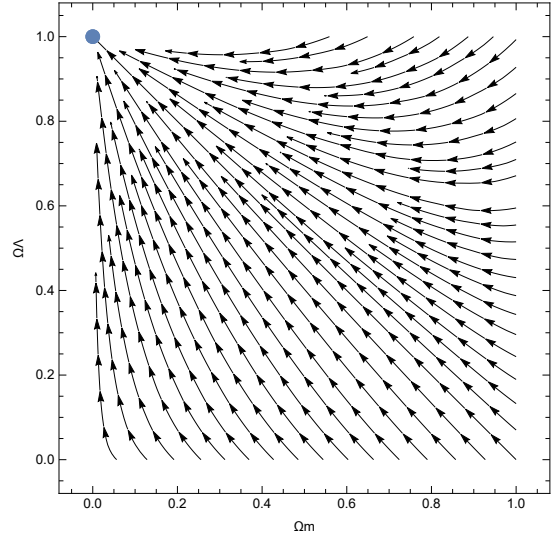}
\caption{Phase space diagram for the dynamical system (\protect\ref{g1}), (%
\protect\ref{g2}). We consider $w_{m}=0$, for $n<1$. The unique attractor is
the de Sitter point $G_{2}$. }
\label{fig:fig1}
\end{figure}

\begin{figure}[ptb]
\centering
\includegraphics[width=0.7\textwidth]{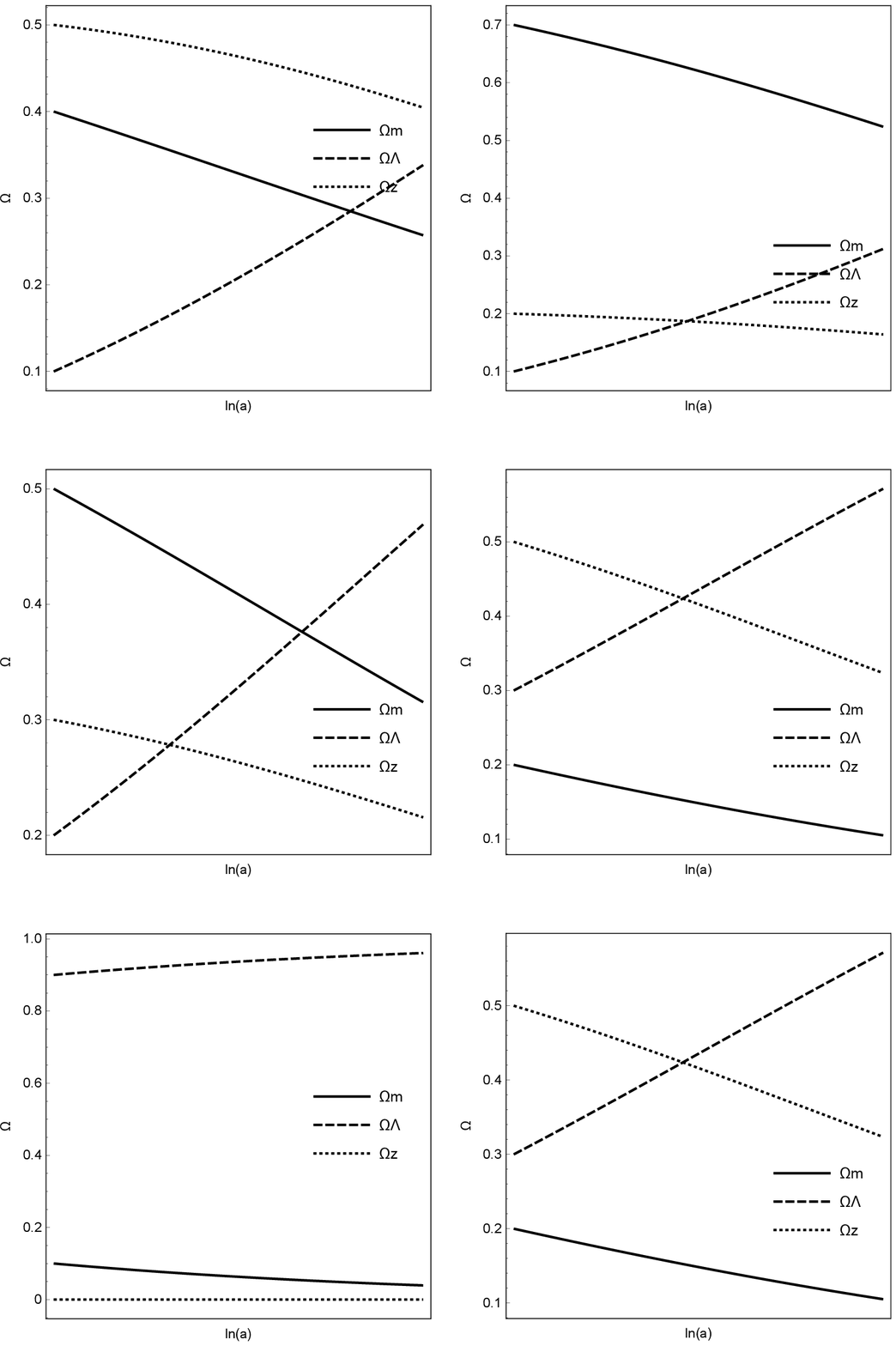}
\caption{Evolution diagrams with time, for various energy densities of the
dynamical system (\protect\ref{g1}), (\protect\ref{g2}). We consider the
initial conditions (a) $\Omega_{m}=0.4,$ $\Omega_{\Lambda}=0.1$ (b) $%
\Omega_{m}=0.7$, $\Omega_{\Lambda}=0.1$ (c) $\Omega_{m}=0.5$, $%
\Omega_{\Lambda}=0.2$ (d) $\Omega_{m}=0.2$, $\Omega _{\Lambda}=0.3$ (e) $%
\Omega_{m}=0.1$, $\Omega_{\Lambda}=0.9$ (f) $\Omega _{m}=0.2$, $%
\Omega_{\Lambda}=0.3$, for $n<1$ and $w_{m}=0$.}
\label{fig:fig3}
\end{figure}

\begin{table}[tbp] \centering%
\caption{Stationary points and physical parameters for the interaction model
G. }%
\begin{tabular}{ccccc}
\hline\hline
\textbf{Point} & $\left( \mathbf{%
\Omega
}_{m}\mathbf{,%
\Omega
}_{\Lambda}\mathbf{,%
\Omega
}_{z}\right) $ & \textbf{Existence} & $\mathbf{w}_{tot}$ & \textbf{%
Acceleration} \\ \hline
$G_{1}$ & $\left( 0,0,-1\right) $ & Always & $-\frac{5}{6}$ & Yes \\ 
$G_{2}$ & $\left( 0,1,0\right) $ & Always & $-1$ & Yes \\ 
$G_{3}$ & $\left( 1,0,0\right) $ & Always & $w_{m}$ & $w_{m}\leq-\frac{1}{3}$
\\ 
$G_{4}$ & $(-\frac{1}{4+6w_{m}},-\frac{5+6w_{m}}{4+6w_{m}},-2-\frac {2}{%
4+6w_{m}})$ & $w_{m}=-\frac{5}{6}$ & $-\frac{5}{6}$ & Yes \\ \hline\hline
\end{tabular}
\label{tab01g}%
\end{table}%
\begin{table}[tbp] \centering%
\caption{Stationary points and stability conditions for the interaction
model G. }%
\begin{tabular}{ccc}
\hline\hline
\textbf{Point} & \textbf{Ei}$\text{\textbf{genvalues}}$ & \textbf{Stability}
\\ \hline
$G_{1}$ & $\{\frac{1}{2},-\frac{3}{2}(1+w_{m})\}$ & Source \\ 
$G_{2}$ & $\{-\frac{1}{2},-\frac{3}{2}(1+w_{m})\}$ & $w_{m}>-1$ \\ 
$G_{3}$ & $\{3(1+w_{m}),\frac{(5+6w_{m})}{2}\}$ & $w_{m}<-1$ \\ 
$G_{4}$ & $\{0,\frac{1}{2}\}$ & Unstable \\ \hline\hline
\end{tabular}
\label{tab01g}%
\end{table}%

\section{Discussion}

\label{sec5}

We performed, for the first time, a detailed study on the dynamics of the
varying vacuum model in a Finsler-Randers geometrical background.
Specifically in the homogeneous and isotropic spatially flat FLRW spacetime
we assumed the existence of an ideal gas fluid source which couples with the
varying vacuum terms. That scenario follows from the interacting models
where interaction in the dark sector has been proposed as a possible
scenario to explain the cosmological observations. For the gravitational
theory, we consider that of Finsler Randers from where a new
geometrodynamical term is introduced and affects the dynamical evolution.

The functional form of varying vacuum model is in generally unknown but a
dominating quadratic term in the Hubble function has been found to be good a
candidate. In this work we consider six different functional forms for the
interaction between the components of the dark sector of the universe.

Models $Q_{A}$, $Q_{B}$ and $Q_{C}$ have been studied in a previous work in
the case of GR \cite{VarVacGP}. In this work we recover the results of the
previous study, that is, the limit of GR relativity is recovered, while
there exists one possible era in the cosmological history which corresponds
to the epoch where only the geometrodynamical term of the FR geometry
contributes.

In addition, we considered three new interaction models, namely $Q_{D}$, $%
Q_{E}$ and $Q_{F}$ which depend also on the geometrodynamical term of FR.
For these three models the limit of GR is recovered while now there is a new
cosmological solution where the geometrodynamical term contributes along the
terms of the dark sector. These new epochs describe accelerated universe. As
far as the stability of these exact solutions are concerned, they can be
stable or unstable, depending on the coupling constants of the models.

Finally $Q_{G}$ is the case without varying vacuum term. In this scenario we
found four critical points which describe the matter dominated era, the de
Sitter universe, the vacuum space and an exact solution which correspond to
a point where all the fluid source contributes in the cosmological evolution.


For model $Q_{A}$\ there are three stationary points, point $A_{1}$\
describes a universe dominated by the Finsler geometrodynamical terms and it
is always a source, points $A_{2}$, $A_{3}$\ describe the limit of GR, where 
$A_{2}$\ describes the de Sitter universe and $A_{3}$\ corresponds to the
solution of GR where the matter source and the cosmological constant term
contribute in the cosmological evolution. Notice that$A_{2}$\ is an
attractor when $w_{m}\geq n-1$\ and $w_{m}$\ is an attractor when $w_{m}<n-1$%
. In the case of a dust fluid, i.e. $w_{m}\,,$\ the de Sitter universe is an
attractor when $n\leq 1$. For model $Q_{B}$\ we determined three stationary
points: $B_{1}$\ describes the Finsler epoch, while $B_{2}$\ corresponds to
the matter dominated era. Point $B_{3}$\ has similar physical properties
with point $A_{3}$, while de Sitter solutions are not provided by the model.
In addition, the Finsler dominated epoch can be an attractor for specific
values of the free parameters, as are the GR solutions of points $B_{2},$~$%
B_{3}$. Model $Q_{C}$\ admits three stationary points, where $C_{1}$\
describes the Finsler epoch and the two points $C_{\pm }$\ describe the
limit of GR which physical properties similar with that of point $A_{3}$. 

Interaction models, $Q_{D}$, $Q_{E}$\ and $Q_{F}$\ provide three stationary
points. Points $D_{1},~D_{2}$\ are the limits of GR, which correspond to the
matter dominated era and the de Sitter universe respectively. Point $D_{3}$\
describes a universe where all the fluid components and the Finsler term
contribute in the cosmological evolution. $~$Points $E_{1},$\ $E_{2}$\ have
physical properties similar to those of points $A_{2},~A_{3}$\ while the
exact solution at point $E_{3}$\ has similarities with the solutions at $%
D_{3}$. Furthermore, the dynamics close to points$F_{1},~F_{2}$is similar
with that of $B_{2},~B_{3}$\ and $F_{3}$\ describes the same epoch with that
of $D_{3}$. 

As far as$Q_{G}$\ model is concerned, the field equations admit four
stationary points. Point $G_{1}$\ describes an unstable exact solution where
the cosmological fluid is described only by the Finsler terms. Point $G_{4}$%
\ is always unstable and is has the same physical properties similar to that
of $D_{3}$. Finally, points $G_{2},~G_{3}$\ are the limit of GR, which
correspond to the de Sitter and matter dominated eras, while one of two
points is a unique attractor. Notice that for $w_{m}>-1$\ \ the attractor is
a de Sitter point.

From the results of this analysis we conclude that the varying vacuum
cosmological scenario in the context of Finsler-Randers geometry can
describe the basic epochs of cosmic history, however there are differences
between the same phenomelogical interaction models in GR. In a future work
we plan to test the performance of the current class of modified gravity
models against the cosmological data. 

\begin{acknowledgments}
GP is supported by the scholarship of the Hellenic Foundation for Research
and Innovation (ELIDEK grant No. 633). SB acknowledges support by the
Research Center for Astronomy of the Academy of Athens in the context of the
program ``Testing general relativity on cosmological scales'' (ref. number
200/872). AP was funded by Agencia Nacional de Investigaci\'{o}n y
Desarrollo - ANID through the program FONDECYT Iniciaci\'{o}n grant no.
11180126. SP was supported by MATRICS (Mathematical Research Impact-Centric
Support Scheme), File No. MTR/2018/000940, given by the Science and
Engineering Research Board (SERB), Govt. of India.
\end{acknowledgments}


\begin{thebibliography}{999}
\bibitem{Riess:1998cb} A.~G.~Riess et al., Astron.\ J.\ 116, 1009 (1998)

\bibitem{Perlmutter:1998np} S.~Perlmutter et al., Astrophys.\ J.\ 517, 565
(1999)

\bibitem{Aghanim:2018eyx} N.~Aghanim et al., arXiv:1807.06209 [astro-ph.CO]

\bibitem{Copeland:2006wr} E.~J.~Copeland, M.~Sami and S.~Tsujikawa, Int.\
J.\ Mod.\ Phys.\ D 15, 1753 (2006)

\bibitem{Peebles:2002gy} P.~J.~E.~Peebles and B.~Ratra, Rev.\ Mod.\ Phys.\
75, 559 (2003)

\bibitem{Padmanabhan:2002ji} T.~Padmanabhan, Phys.\ Rept.\ 380, 235 (2003)

\bibitem{Weinberg:1988cp} S.~Weinberg, Rev.\ Mod.\ Phys.\ 61, 1 (1989)

\bibitem{Sotiriou:2008rp} T.~P.~Sotiriou and V.~Faraoni, Rev.\ Mod.\ Phys.\
82, 451 (2010)

\bibitem{DeFelice:2010aj} A.~De Felice and S.~Tsujikawa, Living Rev.\ Rel.\
13, 3 (2010)

\bibitem{Clifton:2011jh} T.~Clifton, P.~G.~Ferreira, A.~Padilla and
C.~Skordis, Phys.\ Rept.\ 513, 1 (2012)

\bibitem{Capozziello:2011et} S.~Capozziello and M.~De Laurentis, Phys.\
Rept.\ 509, 167 (2011)

\bibitem{Nojiri:2017ncd} S.~Nojiri, S.~D.~Odintsov and V.~K.~Oikonomou,
Phys.\ Rept.\ 692, 1 (2017)

\bibitem{Cai:2015emx} Y.~F.~Cai, S.~Capozziello, M.~De Laurentis and
E.~N.~Saridakis, Rept.\ Prog.\ Phys.\ 79, no. 10, 106901 (2016)

\bibitem{Perelman} C. C. Perelman, Ann. Phys. 416, 168143 (2020)

\bibitem{Minas} E. Minas, E. N. Saridakis, P. Stavrinos and A.
Triantafyllopoulos, Universe 5(3), 74 (2019)

\bibitem{Gibbons} G. W. Gibbons, C. A. R. Herdeiro, C. M. Warnick, M. C.
Werner. Stationary Metrics and Optical Zermelo-Randers-Finsler Geometry.
arXiv:0811.2877 (gr-qc) (2008)

\bibitem{Ikeda} S. Ikeda, E. N. Saridakis and P. C. Stavrinos. Phys. Rev. D
100 (2019)

\bibitem{Kouretsis} A. Kouretsis, M. Stathakopoulos and P. C. Stavrinos.
Phys. Rev. D, 79 (2009)

\bibitem{Hohman} M. Hohmann, C. Pfeifer and N. Voicu. Phys. Rev. D 100 (2019)

\bibitem{hohman2} Manuel Hohmann, Christian Pfeifer, Phys. Rev. D 95, 104021
(2017)

\bibitem{Vacaru} S. Vacaru, Int. J. Mod. Phy. D 21, 1250072 (2012)

\bibitem{Caponio} E. Caponio and G. Stancarone, Int. J. Geom. Meth. Mod.
Phys. Vol 13, No 4, (2016)

\bibitem{Edwards} B. Edwards and A. Kostelecky. Phys. Lett. B, 786 (2018)

\bibitem{Randers} G. Randers, Phys. Rev. 59, 195 (1941)

\bibitem{Stav07} P. Stavrinos, A. Kouretsis and M. Stathakopoulos, Gen. Rel.
Grav. 40, 1403 (2008)

\bibitem{StavFluids} P. C. Stavrinos, Int. J. Theo. Phys. 44, 245 (2005)

\bibitem{15} H. Rund, The Differential Geometry of Finsler Spaces, Springer
(1955)

\bibitem{Bekenstein} J.~D.~Bekenstein, Phys.\ Rev.\ D 48, 3641 (1993)

\bibitem{Mir} R. Miron and M. Anastasiei, `The geometry of Lagrange
spaces:Theory and Applications', Kluwer Academic.Dordrecht (1994)

\bibitem{Bao} D. Bao, S. S. Chern and Z. Shen, `An Introduction to
Riemann-Finsler Geometry', Springer, New York (2000)

\bibitem{13} S. Vacaru, P. Stavrinos, E. Gaburov, D. Gontsa, Clifford and
Riemann - Finsler Structures in Geometric Mechanics and Gravity. Balkan
Press (2005), arXiv: gr-qc/0508023

\bibitem{18} P. Stavrinos, F. Diakogiannis, Grav. Cosmol. 10, 269 (2004)

\bibitem{Asa41} G. S. Asanov, Finsler Geometry, Relativity and Gauge
Theories, Kluwer Academic Publishers Group, Holland (1985)

\bibitem{2} A. Triantafyllopoulos, P.C Stavrinos. Classical and Quantum
Gravity. Vol.35, Is.8, (2018)

\bibitem{4} P. C. Stavrinos, M. Alexiou, Raychaudhuri equation in the
Finsler Randers spacetime and Generalized scalar tensor theories . Int.
Jour. Geom. Meth. Mod. Phys. 15 (03), 1850039, (2018)

\bibitem{5} S. Basilakos, P.Stavrinos. "Cosmological equivalence between the
Finsler-Randers space-time and DGP gravity model". Phys. Rev. D, Vol 87, Is.
4 (2013).

\bibitem{6} S. Basilakos, A. Kouretsis, E. Saridakis, P. C. Stavrinos. Phys.
Rev. D, D 88, Is. 12 (2013). arXiv: 1311.5915 [grqc]. "Resembling dark
energy and modified gravity with Finsler - Randers Cosmology.

\bibitem{7} P. Stavrinos. Weak Gravitational field in Finsler-Randers Space
and Raychchaudhuri Equation. General Relativity and Gravitation Vol.44,
No12, pp.3029. (2012).

\bibitem{8} G. Silva, R.V. Maluf, C.A.S. Almeida. A nonlinear dynamics for
the scalar field in Randers spacetime. Phys. Let. B V. 766, (2017), P.
263-267

\bibitem{9} T. Singh, R. Chaubey and Ashutosh Singh. "Bounce conditions for
FRW models in modified gravity theories". Eur.Phys. Jour. Plus volume 130,
Article number: 31 (2015)

\bibitem{11} Rakesh Raushan R. Chaubey "Finsler Randers cosmology in the
framework of a particle creation mechanism: a dynamical systems
perspective". Eur. Phys.Jour.Plus V. 135, 228 (2020)

\bibitem{12} R. Chaubey, B. Tiwari, Anjani Kumar Shukla and Manoj Kumar.
Finsler Randers Cosmological Models in Modified Gravity Theories Proceedings
of the National Academy of Sciences, India Section A: Physical Sciences V.
89, pages757-768 (2019).

\bibitem{VV} I. L. Shapiro and J. Sol\'{a}, JHEP 0202 (2002) 006; Phys.Lett.
B 475 (2000) 236; J. Sol\'{a}, J. Phys. A 41 (2008) 164066; I. L. Shapiro
and J. Sol\'{a}, Phys. Lett. B 682, 105 (2009); S. Basilakos, MNRAS 395,
2347 (2009); Astron. Astrphys., 508, 575 (2009); S. Basilakos, M. Plionis
and Sol\'{a}, Phys. Rev. D80 (2009) 083511; J. Grande, J. Sol\'{a}, S.
Basilakos and M. Plionis, JCAP 08, 007 (2011); S. Basilakos, J. A. S. Lima
and J. Sol\'{a}, MNRAS 431, 923 (2013); E. L. D. Perico, J. A. S. Lima, S.
Basilakos and J. Sol\'{a}, Phys. Rev. D 88, 063531 (2013); J. Sol\'{a} and
and A. G\'{o}mez-Valent, Int. J. Mod. Phys. D 24, 1541003 (2015); A. G\'{o}%
mez-Valent, J. Sol\'{a} and S. Basilakos, JCAP 01, 004 (2015); V. K.
Oinonomou, S. Pan and R. C. Nunes, Int. J. Mod. Phys. A 32, 1750129 (2017);
S. Pan, Mod. Phys. Lett. A 33, 1850003 (2018); J. Sol\'{a}, J. de Cruz P\'{e}%
rez and A. G\'{o}mez-Valent, EPL 121, 39001 (2018); S. Basilakos, N.
Mavromatos and J. Sol\'{a}, JCAP 12, 025 (2019); Phys. Rev. D 101, 045001
(2020); Phys. Lett. B. 803, 135342 (2020)J. A. S.

\bibitem{mat1} Lima, F. E. Silva and R. C. Santos, Class. Quant. Grav.25,
205006 2008

\bibitem{mat2} G. Steigman, R. C. Santos and J. A. S. Lima, JCAP, 06, 033
2009

\bibitem{mat3} Lima, J. A. S.; Basilakos, S.; Costa, F. E. M., Phys. Rev
D.,86, 103534 2012

\bibitem{matrest} J. A. S. Lima, M. O. Calvao, I. Waga, \textquotedblleft
Cosmology, Thermodynamics and Matter Creation, Frontier Physics, Essays in
Honor of Jayme Tiomno, World Scientific, Singapore (1990), [arXiv:0708.3397];

\bibitem{prigogine} I. Prigogine et al., Gen. Rel. Grav.21, 767 1989

\bibitem{calvao} M. O. Calvao, J. A. S. Lima and I. Waga, Phys. Lett. A162,
223-226 1992

\bibitem{limaGermano} J. A. S. Lima and A. S. M. Germano, Phys. Lett. A 170,
373 1992

\bibitem{zim1} W. Zimdahl and D. Pavon, Gen. Relativ. Grav.,12, 1259 1994

\bibitem{zim2} W. Zimdahl, J. Triginer and D. Pavon Phys. Rev. D.,54, 6101
1996

\bibitem{LimaBasilakos} J. A. S. Lima, S. Basilakos, F. E. M. Costa,
Phys.Rev. D. 86 103534

\bibitem{Parker1968} L. Parker, Phys. Rev. Lett. \textbf{21}, 562 (1968)

\bibitem{Parker1969} L. Parker, Phys. Rev. \textbf{183}, 1057 (1969)

\bibitem{Parker1970} L. Parker, Phys. Rev. \textbf{D3}, 346 (1970)

\bibitem{Parker1977} L. H. Ford and L. Parker, Phys. Rev. \textbf{D16}, 245
(1977)

\bibitem{ZS1972} Ya. B. Zeldovich and A. A. Starobinsky, JETP Lett. \textbf{%
34} 1159 (1972)

\bibitem{ZS1977} a. B. Zeldovich and A. A. Starobinsky, JETP Lett. \textbf{26%
}, 252 (1977)

\bibitem{Grib1974} A. A. Grib, B. A. Levitskii and V. M. Mostepanenko,
Theor. Math. Phys. \textbf{19}, 59 (1974)

\bibitem{Grib1976} A. A. Grib, B. A. Levitskii and V. M. Mostepanenko, Gen.
Rel. Grav. \textbf{7}, 535 (1976)

\bibitem{Grib1994} A. A. Grib, B. A. Levitskii and V. M. Mostepanenko,
Vacuum Quantum Effects in Strong Fields, Friedman Laboratory Publishing,
Saint Petersburg, Russia (1994)

\bibitem{Sch1939} E. Schr\"{o}dinger, Physica \textbf{6}, 899 (1939)

\bibitem{Steigman:2008bc} G.~Steigman, R.~C.~Santos and J.~A.~S.~Lima, 
JCAP \textbf{06}, 033 (2009) 

\bibitem{Lima:2009ic} J.~A.~S.~Lima, J.~F.~Jesus and F.~A.~Oliveira, 
JCAP \textbf{11}, 027 (2010) 

\bibitem{Lima:2014qpa} J.~A.~S.~Lima, L.~L.~Graef, D.~Pavon and
S.~Basilakos, 
JCAP \textbf{10}, 042 (2014) 

\bibitem{Nunes:2015rea} R.~C.~Nunes and D.~Pav\u{A}\l n, 
Phys. Rev. D \textbf{91}, 063526 (2015) 

\bibitem{Pigozzo:2015swa} C.~Pigozzo, S.~Carneiro, J.~S.~Alcaniz,
H.~A.~Borges and J.~C.~Fabris, 
JCAP \textbf{05}, 022 (2016) 

\bibitem{Amendola:1999er} L.~Amendola, Phys.\ Rev.\ D 62, 043511 (2000)

\bibitem{Amendola:2003eq} L.~Amendola and C.~Quercellini, Phys.\ Rev.\ D 68,
023514 (2003)

\bibitem{Cai:2004dk} R.~G.~Cai and A.~Wang, JCAP 0503, 002 (2005)

\bibitem{Pavon:2005yx} D.~Pav\'{o}n and W.~Zimdahl, Phys.\ Lett.\ B 628, 206
(2005)

\bibitem{delCampo:2008sr} S.~del Campo, R.~Herrera and D.~Pav\'{o}n, Phys.\
Rev.\ D 78, 021302 (2008)

\bibitem{delCampo:2008jx} S.~del Campo, R.~Herrera and D.~Pav\'{o}n, JCAP
0901, 020 (2009)

\bibitem{Wetterich:1994bg} C.~Wetterich, Astron.\ Astrophys.\ 301, 321 (1995)

\bibitem{Barrow:2006hia} J.~D.~Barrow and T.~Clifton, Phys.\ Rev.\ D 73,
103520 (2006)

\bibitem{Amendola:2006dg} L.~Amendola, G.~Camargo Campos and R.~Rosenfeld,
Phys.\ Rev.\ D 75, 083506 (2007)

\bibitem{Pavon:2007gt} D.~Pav\'{o}n and B.~Wang, Gen.\ Rel.\ Grav.\ 41, 1
(2009)

\bibitem{Chimento:2009hj} L.~P.~Chimento, Phys.\ Rev.\ D 81, 043525 (2010)

\bibitem{Arevalo:2011hh} F.~Arevalo, A.~P.~R.~Bacalhau and W.~Zimdahl,
Class.\ Quant.\ Grav.\ 29, 235001 (2012)

\bibitem{Yang:2018xlt} W.~Yang, S.~Pan, R.~Herrera and S.~Chakraborty, Phys.
Rev. D 98, no.4, 043517 (2018)

\bibitem{an001} A. Paliathanasis, S. Pan and W. Yang, Int. J. Mod. Phys. D
28, 1950161 (2019)

\bibitem{Tsiapi:2018she} P.~Tsiapi and S.~Basilakos, MNRAS 485 (2019)

\bibitem{Yang:2018qec} W.~Yang, N.~Banerjee, A.~Paliathanasis and S.~Pan,
Phys. Dark Univ. 26, 100383 (2019)

\bibitem{Pan:2020mst} S.~Pan, J.~de Haro, W.~Yang and J.~Amor\'{o}s,
[arXiv:2001.09885 [gr-qc]]

\bibitem{Salvateli} V. Salvatelli, N. Said, M. Bruni, A. Melchiorri and D.
Wands, Phys. Rev. Lett. 113, 181301 (2014)

\bibitem{QproptorhoL} R.~Murgia, S.~Gariazzo and N.~Fornengo, JCAP 1604
(2016)

\bibitem{Nunes:2016dlj} R.~C.~Nunes, S.~Pan and E.~N.~Saridakis, Phys. Rev.
D 94, no.2, 023508 (2016)

\bibitem{Pan:2016ngu} S.~Pan and G.~Sharov, MNRAS 472, no.4, 4736 (2017)

\bibitem{Sharov:2017iue} G.~S.~Sharov, S.~Bhattacharya, S.~Pan, R.~C.~Nunes
and S.~Chakraborty, MNRAS 466, no.3, 3497 (2017)

\bibitem{Yang:2017yme} W.~Yang, N.~Banerjee and S.~Pan, Phys. Rev. D 95,
no.12, 123527 (2017)

\bibitem{Yang:2017ccc} W.~Yang, S.~Pan and D.~F.~Mota, Phys. Rev. D 96,
no.12, 123508 (2017)

\bibitem{Yang:2017zjs} W.~Yang, S.~Pan and J.~D.~Barrow, Phys. Rev. D 97,
no.4, 043529 (2018)

\bibitem{Pan:2017ent} S.~Pan, A.~Mukherjee and N.~Banerjee, MNRAS 477, no.1,
1189 (2018)

\bibitem{Yang:2018euj} W.~Yang, S.~Pan, E.~Di Valentino, R.~C.~Nunes,
S.~Vagnozzi and D.~F.~Mota, JCAP 09, 019 (2018)

\bibitem{in1} D. Begue, C. Stahl and S.-S. Xue, Nucl. Phys. B 940, 312 (2019)

\bibitem{in2} M. Szydlowski, T. Stachowiak and R. Wojtak, Phys. Rev. D 73,
063516 (2006)

\bibitem{in3} W. Yang, S. Pan and A. Paliathanasis, MNRAS 482, 1007 (2019)

\bibitem{Pan:2019jqh} S.~Pan, W.~Yang, C.~Singha and E.~N.~Saridakis, Phys.
Rev. D 100, no.8, 083539 (2019)

\bibitem{Pan:2019gop} S.~Pan, W.~Yang, E.~Di Valentino, E.~N.~Saridakis and
S.~Chakraborty, Phys. Rev. D 100, no.10, 103520 (2019)

\bibitem{in4} S. Pan, W. Yang and A. Paliathanasis, MNRAS 493, 3114 (2020)

\bibitem{in5} W. Yang, S. Pan, R. C. Nunes and D. F.\ Mota, JCAP 04, 008
(2020)

\bibitem{DiValentino:2019ffd} E.~Di Valentino, A.~Melchiorri, O.~Mena and
S.~Vagnozzi, [arXiv:1908.04281 [astro-ph.CO]]

\bibitem{DiValentino:2019jae} E.~Di Valentino, A.~Melchiorri, O.~Mena and
S.~Vagnozzi, Phys. Rev. D 101, no.6, 063502 (2020)

\bibitem{Pan:2020zza} S.~Pan, G.~S.~Sharov and W.~Yang, [arXiv:2001.03120
[astro-ph.CO]]

\bibitem{Prigogine-inf} I. Prigogine, J. Geheniau, E. Gunzig, P. Nardone,
Gen. Rel. Grav. 21, 767 (1989)

\bibitem{Abramo:1996ip} L.~R.~W.~Abramo and J.~A.~S.~Lima, Class.\ Quant.\
Grav.\ 13, 2953 (1996)

\bibitem{Gunzig:1997tk} E.~Gunzig, R.~Maartens and A.~V.~Nesteruk, Class.\
Quant.\ Grav.\ 15, 923 (1998)

\bibitem{Zimdahl:1999tn} W.~Zimdahl, Phys.\ Rev.\ D 61, 083511 (2000)

\bibitem{RCN16} R. C. Nunes, Int. J. Mod. Phys. D 25, 1650067 (2016)

\bibitem{deHaro:2015hdp} J.~de Haro and S.~Pan, Class. Quant. Grav. 33,
no.16, 165007 (2016)

\bibitem{sp1} S. Pan, J. Haro, A. Paliathanasis and R. J. Slagter, MNRAS
460, 1445 (2016)

\bibitem{Nunes:2016aup} R.~C.~Nunes and S.~Pan, MNRAS 459, no.1, 673 (2016)

\bibitem{sp2} A. Paliathanasis, J. D.\ Barrow and S. Pan, Phys. Rev. D 95,
103516 (2017)

\bibitem{VarVacGP} G. Papagiannopoulos, P. Tsiapi, S. Basilakos and A.
Paliathanasis, Eur. Phys. J. C 80, 55 (2020)

\bibitem{finslerGP} G. Papagiannopoulos, S. Basilakos, A Paliathanasis and
P. C. Stavrinos, Class. Quant. Grav. 34, 22 (2017)

\bibitem{con1} E. J. Copeland, A. R. Liddle and D. Wands, Phys. Rev. D. 57,
4686 (1998)

\bibitem{con2} C. R. Fadragas and G. Leon, Class. Quant. Grav. 31, 195011
(2014)

\bibitem{wiggins} S. Wiggins, Introduction to applied nonlinear dynamical
systems and chaos, Springer-Verlag, New\ York, (1990)

\bibitem{aetherGP} A. Paliathanasis, G. Papagiannopoulos, S. Basilakos and
J. D. Barrow, Eur. Phys. J. C 79, 723 (2019)

\bibitem{Quintessence} S. Basilakos, G. Leon, G. Papagiannopoulos and E. N.
Saridakis, Phys. Rev. D 100, 043524 (2019)

\bibitem{BransDickeGP} G. Papagiannopoulos, J. D. Barrow, S. Basilakos, A.
Giacomini and A. Paliathanasis, Phys. Rev. D 95, 024021 (2017)

\bibitem{dyn2} G. Leon and E. N. Saridakis, JCAP 1504, 031 (2015)

\bibitem{dyn3} G. Leon, Int. J. Mod. Phys. E 20, 19 (2011)

\bibitem{dyn4} T. Gonzales, G. Leon and I. Quiros, Class. Quant. Grav. 23,
3165 (2006)

\bibitem{dyn5} A. Giacomini,\ S. Jamal, G. Leon, A. Paliathanasis and J.\
Saveedra, Phys.\ Rev. D 95, 124060 (2017)

\bibitem{dyn6} G. Chee and Y. Guo, Class. Quantum Grav. 29, 235022 (2012)
[Corrigendum: Class. Quant. Grav. 33, 209501 (2016)]

\bibitem{dyn7} S. Mishra and S.\ Chakraborty, Eur. Phys. J. C 79, 328 (2019)

\bibitem{dyn8} H. Farajollahi and A. Salehi, JCAP 07, 036 (2011)

\bibitem{dyn9} A. Paliathanasis, Phys. Rev. D 101, 064008 (2020)

\bibitem{pano1} G. Panotopoulos, A. Rincon, N. Videla and G. Otalora, Eur.
Phys. J. C 80, 286 (2020)

\bibitem{Kerachian} M. Kerachian, G. Acquaviva, G.L. Gerakopoulos, Phys.
Rev. D 101, 043535 (2020)
\end{thebibliography}
\end{document}